\documentclass[
  reprint, prx,
  superscriptaddress,
  amsmath,amssymb,
  aps,floatfix,
  showkeys]{revtex4-2}
\usepackage[dvipsnames]{xcolor}
\usepackage{tcolorbox}
\usepackage{wrapfig}
\usepackage{floatrow}
\usepackage[english]{babel}
\usepackage{graphicx}
\usepackage{nccmath}
\usepackage{float}

\usepackage{cleveref}
\crefname{equation}{Eq.}{Eqs.}
\crefrangelabelformat{equation}{(#3#1#4--#5#2#6)}

\usepackage{amsmath}

\usepackage{mathrsfs}
\usepackage{textcomp, gensymb}
\usepackage{nccmath}
\usepackage{mathtools} 
\newcommand{\lb}{\langle}
\newcommand{\rb}{\rangle}
\newcommand{\comm}[1]{\textcolor{black}{#1}}

\global\long\def\Cov{\mathrm{Cov}}
\global\long\def\CV{\mathrm{CV}}

\global\long\def\Var{\mathrm{Var}}

\newcommand{\CC}{C\nolinebreak\hspace{-.05em}\raisebox{.4ex}{\tiny\bf +}\nolinebreak\hspace{-.10em}\raisebox{.4ex}{\tiny\bf +}}

\usepackage{lipsum}

\begin{document}

\preprint{APS/123-QED}

\title{Efficiency-fluctuation trade-offs in biomolecular assembly processes}

\author{Brayden Kell}
\affiliation{Department of Physics, University of Toronto, Toronto, ON M5S 1A7, Canada}
\affiliation{Department of Chemical \& Physical Sciences, University of Toronto, Mississauga, ON L5L 1C6, Canada}
\author{Andreas Hilfinger}
\email{andreas.hilfinger@utoronto.ca}
\affiliation{Department of Physics, University of Toronto, Toronto, ON M5S 1A7, Canada}
\affiliation{Department of Chemical \& Physical Sciences, University of Toronto, Mississauga, ON L5L 1C6, Canada}

\begin{abstract}
Stochastic fluctuations of molecular abundances are a ubiquitous feature of cellular processes and lead to significant cell-to-cell variability.
Recent theoretical work established
lower bounds for stochastic fluctuations in cells for broad classes of cellular processes 
by analyzing the dynamics of reaction motifs that are embedded within %
\comm{a larger network with arbitrary interactions and dynamics}. For example, %
a class of generalized assembly processes in which two co-regulated subunits irreversibly form a complex was shown to exhibit an unavoidable trade-off between assembly efficiency and subunit fluctuations:
Regardless of rate constants and details of feedback control,  subunit fluctuations were shown to diverge as the assembly efficiency approaches 100\%. In contrast, other work has reported how efficient assembly processes work as stochastic noise filters or can achieve robust adaptation through integral control. While all of these results are technically correct their seemingly contradictory conclusions raise the question of how broadly applicable the previously reported efficiency-fluctuation trade-off is. Here, we show that a much broader class of assembly processes than previously considered is subject to an efficiency-fluctuation trade-off which diverges in the high efficiency regime. 
We find the proposed noise filtering property of efficient assembly processes corresponds to a singular limit of this 
class of systems. Additionally, we show that combining feedback control with \emph{distinct} subunit synthesis rates is \comm{a necessary condition to overcome the generalized efficiency-fluctuation trade-off}. Through numerical examples, we show that biomolecular integral controllers are one of several realizations of such control. %
\comm{How small a change to joint subunit control is sufficient to avoid diverging fluctuations in the high-efficiency limit remains an open question.}
\end{abstract}


\maketitle

\section{Introduction}
Molecular abundances in cells fluctuate significantly due to the probabilistic nature of individual biochemical reactions~\cite{Spudich1976, Elowitz2002, Blake2003}. \comm{Such fluctuations have phenotypic consequences~\cite{Ozbudak2002, Raj2008}, but analyzing how stochastic fluctuations are generated and transmitted in cells is challenging due to the complexity and incomplete characterization of intracellular interactions.} \comm{Despite the complexity of biochemical reaction networks,} recent theoretical work has characterized such stochastic fluctuations by deriving lower bounds that highlight general trade-offs for broad classes of %
networks~\cite{Lestas2010,Hilfinger2016a,yan2019kinetic}. \comm{Specifically, fluctuation control limits have been derived for reaction modules embedded in arbitrary networks~\cite{Hilfinger2016a}.}

\comm{An important class of biochemical reaction modules are those where two or more components self-assemble into a complex. Such assembly processes are a ubiquitous feature of all biological cells. For example, a multi-subunit complex is involved in transcription initiation in eukaryotes, archaea, and bacteria alike~\cite{Alberts2002}. Assembly also underpins the synthesis of crucial cellular components such as ribosomal complexes, transmembrane transporters~\cite{Li2014}, or luciferases that allow bacteria to emit light~\cite{Baldwin1995,Schaefer1996}.}

Previous work analyzed generalized assembly processes in which two molecular subunits share an arbitrary synthesis rate, undergo first order degradation with a shared timescale, and react irreversibly to form a complex through mass action kinetics. All possible systems that share those reactions must exhibit an unavoidable trade-off between assembly efficiency 
and the lowest possible level of subunit fluctuations~\cite{Hilfinger2016a}. In particular, regardless of the nature of feedback control or the values of rate constants, the subunit fluctuations of such systems must diverge to infinity as the assembly efficiency approaches 100\%. 
\comm{The reported efficiency-fluctuation trade-off thus predicts large subunit fluctuations as a necessary consequence of assembly processes that have evolved to minimize the costly synthesis of proteins, which constitutes a significant amount of the total energy expenditure of cells~\cite{Buttgereit1995, Russell1995, Li2014}.}

However, other work that considered fluctuations in cellular assembly processes has reported seemingly contradictory conclusions. For example, a perfectly efficient complexing motif has been reported to act as a stochastic noise filter \cite{Laurenti2018} and antithetic integral feedback control modules~\cite{Briat2016, Aoki2019}, which rely on perfectly efficient complexing to achieve perfect adaptation of average abundances, have been shown to achieve finite subunit fluctuations~\cite{Briat2016}.%

There is no technical contradiction between the above results because the  %
class of systems that must obey the reported trade-off~\cite{Hilfinger2016a} excludes the types of processes that exhibit finite fluctuations and 100\% assembly efficiency \cite{Laurenti2018,Briat2016}. However, if important types of assembly processes are excluded from the previously considered class of systems, it calls into question the broad conclusions of the previously reported ``general'' trade-off~\cite{Hilfinger2016a}.

Here, we revisit this efficiency-fluctuation trade-off and analyze the effect of changing the assumptions that define the class of generalized assembly processes considered previously. In particular, we consider the effect of allowing for i) simultaneous subunit synthesis in a single reaction; ii) unequal subunit degradation rates; iii) reversibility of complex formation; iv) non mass-action kinetics for complex formation; v) complexes with more than one subunit copy; and vi) separate open-loop control for each of the subunits.
We find that, as long as the control of subunit synthesis is shared, 
the general trade-off in which subunit fluctuations must diverge as the assembly efficiency approaches 100\% generalizes to all such systems.
Particularly, we show that the finite fluctuations reported for systems with strictly simultaneous subunit synthesis~\cite{Laurenti2018} correspond to a singular limit of this class of biochemical reaction networks.

In contrast, systems that combine feedback control with \emph{distinct} subunit synthesis rates can overcome the efficiency-fluctuation trade-off and exhibit finite fluctuations even for an assembly efficiency of 100\%. We show that antithetic integral feedback~\cite{Briat2016} is one of several realizations of such %
control that avoid diverging fluctuations in the high-efficiency regime. 

\comm{Mathematically, our work is based on modeling %
biochemical reactions as continuous-time Markov processes with a discrete, but infinite, state-space. 
In particular, we analyze incompletely specified systems with a set of intracellular species of interest  $X_1,\ldots, X_n$ whose abundance is denoted by $\boldsymbol{x} = (x_0,\ldots,x_n)\in \mathbb{Z}_{\geq 0}^n$.
We specify only the $m$ reactions in the systems that change the abundance of the components of interest
\begin{equation}\label{eq: jump process}
    \boldsymbol{x} \overset{\mathcal{R}_k(\boldsymbol{u},\boldsymbol{x})}{\xrightarrow{\hspace*{1.15cm}}} \boldsymbol{x} + \boldsymbol{d}_k\ , \hspace{1em} k = 1, \ \ldots \ m \ ,
\end{equation}
where $d_{ik}\in \mathbb{Z}$ is the jump size of $X_i$ in the $k^\text{th}$ reaction, and $\mathcal{R}_k(\boldsymbol{x}, \boldsymbol{u})$ denotes the propensity of the $k^\text{th}$ reaction, which depends on the current abundances of the specified components $\boldsymbol{x}$ and arbitrarily many unspecified components $\boldsymbol{u} \coloneqq u_1,u_2,\ldots$. Note that while the specified components undergo Markovian transitions, we allow the unspecified components $\boldsymbol{u}$ to follow arbitrary, potentially non-Markovian, dynamics.~\cite{Hilfinger2016a}}

Throughout, we quantify stochastic fluctuations \comm{of the species of interest} in terms of their coefficient of variation
\begin{equation}\label{eq: CV}
    \CV_i \coloneqq \frac{\sqrt{\mathrm{Var}(x_i)}}{\left\lb x_i \right\rb}\ , 
\end{equation}
where $\langle x_i \rangle$ and $\mathrm{Var}(x_i)$ denote the stationary-state mean and variance for the abundance of component $X_i$. \comm{Using stationary moment invariants between pairs of specified components~\cite{Hilfinger2016a}, allows us to derive lower bounds on fluctuations for specified components within broad classes of assembly processes. Throughout, our results apply to ergodic systems for which time-averages of the specified components within individual systems are equivalent to stationary-state ensemble averages. See Appendix Sects.~\ref{sect: dynamics of incompletely specified systems},~\ref{sect: general relations} and Ref.~\cite{Hilfinger2016a} for details of the approach.}

\section{Results}

\subsection{Systems with shared subunit synthesis rates}\label{sect: equal synthesis}
\begin{figure*}[ht]
    \centering
    \includegraphics[width = 0.85\linewidth]{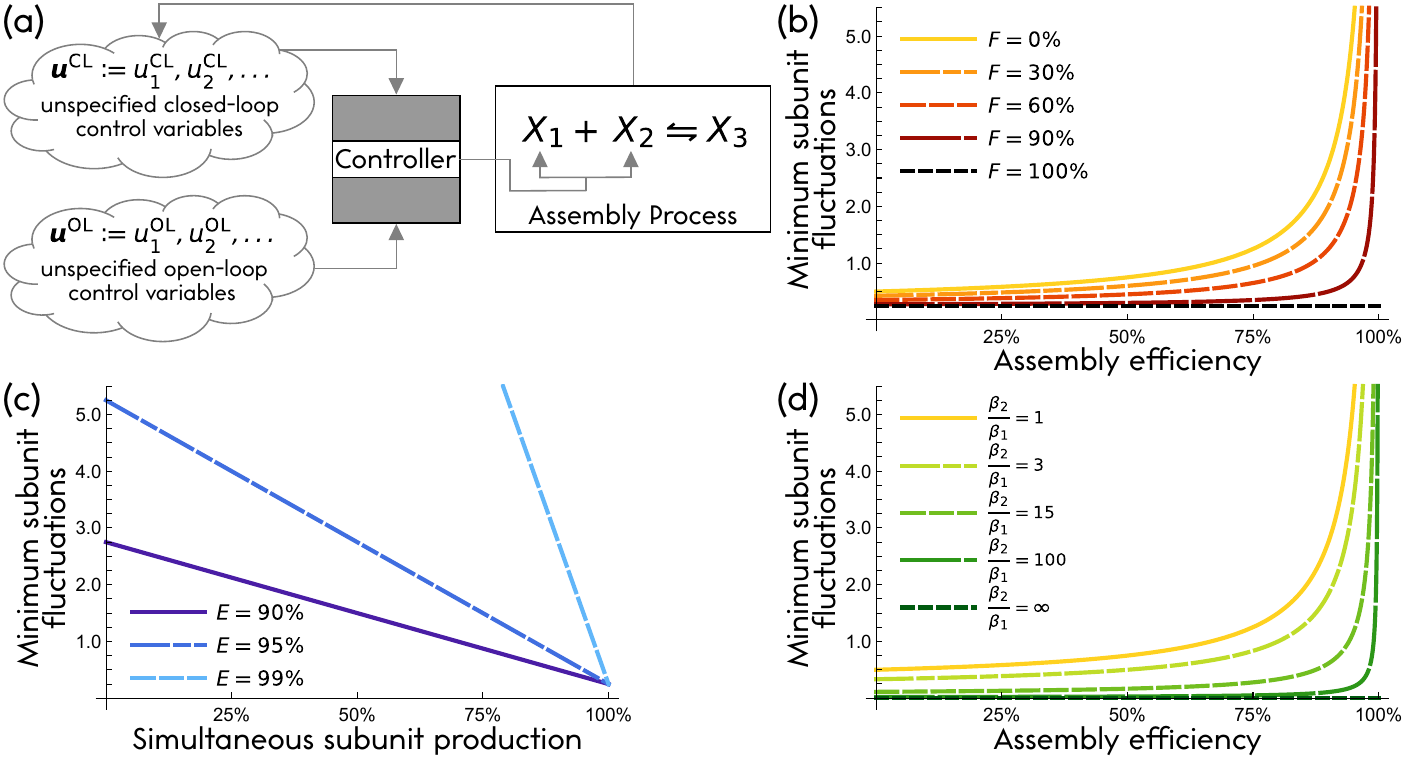}
    \caption{\textit{Assembly processes under arbitrary single-input control obey a diverging efficiency-fluctuation trade-off.} (a) We consider all possible assembly processes in which two subunits $X_1,X_2$ form a complex $X_3$ with 1:1 stoichiometry and the control of subunit synthesis rates is the same for each subunit, as defined in Eqs.~(\ref{eq: 2016 with co-synthesis - subunit dynamics},~\ref{eq: 2016 with co-synthesis - complexing dynamics}). (b) The minimum subunit fluctuations increase monotonically as a function of assembly efficiency $E$, see \cref{eq: beta asymmetry}. The value of $F$ denotes the fraction of subunits that enter the process simultaneously, see \cref{eq: F defn}. Plotted is the lower bound for different values of $F$ and symmetric subunit degradation, i.e., $\beta_1 = \beta_2$. Subunit fluctuations are quantified relative to the Poisson fluctuations subunits would exhibit with constant synthesis rates in the absence of complex formation.
    Increasing $F$ decreases the minimum fluctuations for a given efficiency value, but for all $F < 100\%$ subunit fluctuations diverge as the assembly efficiency approaches $100\%$. (c) The lower bound of~\cref{eq: beta asymmetry} as a function of the simultaneous subunit production ratio $F$ for different efficiency values and $\beta_1 = \beta_2$. (d) The lower bound of~\cref{eq: beta asymmetry} for different degradation rate asymmetries $\beta_2/\beta_1 \geq 1$ while keeping $F = 0$. Increasing $\beta_2/\beta_1$ decreases the minimum fluctuations for a given efficiency value. For any finite value of $\beta_2/\beta_1$ subunit fluctuations must diverge as the assembly efficiency approaches $100\%$. %
    }
    \label{fig: single-input}
\end{figure*}

We first consider the following class of assembly processes in which the %
the subunits $X_1, X_2$ %
\comm{are synthesized and degraded according to the following transitions:}
\begin{equation}\label{eq: 2016 with co-synthesis - subunit dynamics}
    \begin{gathered}
            {x_1 \overset{r_I}{\xrightarrow{\hspace*{1.1cm}}} x_1 + 1 \hspace{0.47cm} x_2  \overset{r_I}{\xrightarrow{\hspace*{1.1cm}}} x_2 + 1} \vspace{-.3em}\\ \text{\emph{\footnotesize{individual subunit synthesis}}}\\
         (x_1,x_2) \overset{r_S}{\xrightarrow{\hspace*{1.1cm}}} (x_1+1,x_2+1)\vspace{-.3em}\\
         \text{\emph{\footnotesize{simultaneous subunit synthesis}}}\\
        x_1 \overset{\beta_1 x_1}{\xrightarrow{\hspace*{1.1cm}}} x_1-1 \hspace{0.47cm} x_2 \overset{\beta_2 x_2}{\xrightarrow{\hspace*{1.1cm}}} x_2-1 \vspace{-.3em}\\
        \text{\emph{\footnotesize{subunit degradation}}}\\
    \end{gathered}
\end{equation}
where the synthesis rates
\begin{eqnarray*}
r_I  &= r_I(\boldsymbol{u}^\mathrm{OL},\boldsymbol{u}^\mathrm{CL}, \boldsymbol{x})\\
r_S &=r_S(\boldsymbol{u}^\mathrm{OL},\boldsymbol{u}^\mathrm{CL}, \boldsymbol{x})
\end{eqnarray*}
are left completely unspecified and thus allow for arbitrary types of feedback, extrinsic noise, or open-loop control. In the functional dependencies we distinguish between the variables involved in the assembly process itself $\boldsymbol{x}\coloneqq (x_1,x_2,x_3)$, an arbitrary number of open-loop control variables $\boldsymbol{u}^\mathrm{OL} \coloneqq (u_1^\mathrm{OL}, u_2^\mathrm{OL},\ldots)$, which are causally unaffected by $\boldsymbol{x}$, and closed-loop control variables $\boldsymbol{u}^\mathrm{CL}\coloneqq (u_1^\mathrm{CL}, u_2^\mathrm{CL},\ldots)$, which are directly or indirectly affected by $\boldsymbol{x}$, see Fig.~\ref{fig: single-input}(a). 

The assembly dynamics of the complex $X_3$ is in turn given by
\begin{equation}\label{eq: 2016 with co-synthesis - complexing dynamics}
    \begin{gathered}
        (x_1, x_2 ,x_3) \overset{r_A}{\xrightarrow{\hspace*{1.1cm}}} (x_1 - 1, x_2 - 1, x_3 + 1)\vspace{-.3em} \\
        \text{\emph{\footnotesize{complex assembly}}} \\
        (x_1, x_2 ,x_3) \overset{r_D}{\xrightarrow{\hspace*{1.1cm}}} (x_1 + 1, x_2 + 1, x_3 - 1)\vspace{-.3em} \\
        \text{\emph{\footnotesize{complex dissociation}}} \\
        x_3 \overset{\beta_3 x_3}{\xrightarrow{\hspace*{1.1cm}}} x_3-1\vspace{-.3em} \\
        \text{\emph{\footnotesize{complex degradation}}}
    \end{gathered}
\end{equation}
where the association and dissociation rates 
\begin{eqnarray*}
r_A  &= r_A(\boldsymbol{u}^\mathrm{OL},\boldsymbol{u}^\mathrm{CL}, \boldsymbol{x})\\
r_D &=r_D(\boldsymbol{u}^\mathrm{OL},\boldsymbol{u}^\mathrm{CL}, \boldsymbol{x})
\end{eqnarray*}
are left completely unspecified and can depend in arbitrary ways on both the specified components $\boldsymbol{x}$ and the %
unspecified components \comm{$\boldsymbol{u}^\mathrm{OL},\boldsymbol{u}^\mathrm{CL}$}. 

Consistent with previous work~\cite{Hilfinger2016a}, we define the assembly efficiency as the fraction of subunits which form a complex (at least once) before being degraded
\begin{equation}\label{eq: E defn}
    E \coloneqq \frac{\langle r_A \rangle}{\langle r_A \rangle + \beta_1 \langle x_1 \rangle}\ .
\end{equation}
The quantity $E$ is an efficiency measure because $1-E$ \comm{is the fraction of subunits that is degraded before forming a complex. The synthesis of such units is thus guaranteed to be wasted. Other kinetic details, such as rapid complex degradation, can make the subunit synthesis even more wasteful but $E$ can be intuitively understood as a lower bound for the efficiency of subunit utilization.}

Note that while the production of $X_1$ and $X_2$ is controlled by %
\comm{shared rates}, the %
\comm{control functions} are not assumed to be symmetric with respect to control, i.e., $r_I, r_S, r_A, r_D$ are permitted to depend asymmetrically on $X_1$ and $X_2$ -- both directly and through the closed-loop control variables $\boldsymbol{u}^\mathrm{CL}$ -- which can lead to asymmetric subunit fluctuations. Furthermore, the degradation rates $\beta_1, \beta_2$ are allowed to be unequal which leads to asymmetric subunit averages. However, even for such asymmetric subunit dynamics, both subunits exhibit the same assembly efficiency, because flux balance at stationarity implies $\beta_1 \langle x_1  \rangle = \beta_2\langle x_2 \rangle$ for the dynamics defined by Eqs.~(\ref{eq: 2016 with co-synthesis - subunit dynamics},~\ref{eq: 2016 with co-synthesis - complexing dynamics}), see Appendix \cref{eq: R_i}.

The class of assembly processes defined in Eqs.~(\ref{eq: 2016 with co-synthesis - subunit dynamics},~\ref{eq: 2016 with co-synthesis - complexing dynamics}) and illustrated in Fig.~\ref{fig: single-input}(a) is much broader than the one previously discussed in Ref.~\cite{Hilfinger2016a}: i) We allow for unequal subunit degradation rates $\beta_1 \neq \beta_2$. ii) We generalize the complex formation from mass-action kinetics to arbitrary association rates and allow for the dissociation of the complex, likewise with an arbitrary rate. iii) The synthesis of subunits is not assumed to be coupled only through a shared rate governing individual subunit synthesis, but we allow for simultaneous de novo synthesis of both subunits in a shared reaction. %

Although strictly simultaneous de novo synthesis of subunits is arguably unrealistic for most cellular processes, the reported behavior of finite subunit fluctuations despite 100\% efficiency in such processes \cite{Laurenti2018} highlights an important technical question: namely, whether the previously reported bound~\cite{Hilfinger2016a} depends on making very specific assumptions such that even small deviations from those assumptions could lead to drastically different behavior. %

To relate the generalized class of systems Eqs.~(\ref{eq: 2016 with co-synthesis - subunit dynamics},~\ref{eq: 2016 with co-synthesis - complexing dynamics}) to the previously considered special case of $r_S=0$ and \mbox{$r_D=0$} we introduce the simultaneous subunit production \comm{ratio}
\begin{equation}\label{eq: F defn}
    F \coloneqq \frac{\langle r_S \rangle + \langle r_D \rangle}{\langle r_I \rangle + \langle r_S \rangle + \langle r_D \rangle}\ .
\end{equation}
Here, and throughout, the term ``simultaneous production'' refers to not just 
de novo synthesis events but includes the simultaneous appearance of subunits when a complex dissociates.

For all possible assembly processes that contain the reactions defined in Eqs.~(\ref{eq: 2016 with co-synthesis - subunit dynamics},~\ref{eq: 2016 with co-synthesis - complexing dynamics}) we show that the subunit fluctuations are constrained by the following bound
\begin{equation}\label{eq: beta asymmetry}
\frac{1}{2}\left(\frac{\CV_1^2}{1/\langle x_1 \rangle} + \frac{\CV_2^2}{1/\langle x_2 \rangle}\right) \geq \frac{1}{3+\mfrac{\beta_2}{\beta_1}}\left(1+\frac{1-F}{1-E}\right)\ ,
\end{equation}
where $\beta_2 \geq \beta_1$ is assumed without loss of generality. This inequality is proven mathematically in Appendix~\ref{sect: beta appendix} using exact stationary moment invariants and basic mathematical inequalities without any approximations. 

The left hand side of \cref{eq: beta asymmetry} corresponds to the average subunit fluctuations relative to the Poisson fluctuations that subunits would exhibit with constant synthesis rates in the absence of complex formation. Note that \cref{eq: beta asymmetry} reduces to the previously reported bound \cite{Hilfinger2016a} for $F=0$ and $\beta_1 = \beta_2$. For any value of $F<100\%$, the bound diverges as the efficiency approaches 100\%, see Fig.~\ref{fig: single-input}(b). For any $E<100\%$, increasing $F$ linearly decreases the severity of the trade-off between assembly efficiency and minimum subunit fluctuations, see Fig.~\ref{fig: single-input}(c). %

When individual subunit synthesis events \emph{never} occur such that $F$ is \emph{exactly} 100\%, the divergence disappears and the right hand side of \cref{eq: beta asymmetry} 
becomes independent of the assembly efficiency $E$.
This explains the observation that systems with exclusively simultaneous subunit synthesis can achieve finite subunit fluctuations at 100\% efficiency~\cite{Laurenti2018}. However, the bound of \cref{eq: beta asymmetry} also highlights the singular nature of this noise suppression: even an %
arbitrarily infrequent appearance of individual subunits
leads to infinite subunit fluctuations as the efficiency approaches 100\%. 
The previously considered behaviour of $F=0$ thus generalizes to the interval $F\in[0,1)$ and excludes only the edge case of $F=100\%$. \comm{For completeness we address the singular case of $E=F=1$ in Appendix \ref{sect: E=F=1} where we show that the dynamics become non-ergodic in general and map to a single-variable problem where noise filtering~\cite{Laurenti2018} can be understood as a consequence of effective quadratic degradation in the case of mass action assembly. %
}

The singular behaviour of $F=100\%$ highlights that setting reaction rates strictly to zero versus infinitesimally small ones can drastically change the conclusions drawn from mathematical analyses of reaction networks. We previously reported similar observations for the minimum fluctuations achievable in antithetic integral feedback systems~\cite{Kell2023}. Understanding biological systems thus requires analyzing what happens when real systems deviate from idealized assumptions. %
For example, this is why we extend the previously considered class of assembly processes to allow for complex dissociation in \cref{eq: 2016 with co-synthesis - complexing dynamics}: Even if the dissociation reaction might be rare in a given assembly process, it is energetically impossible for the dissociation reaction to \emph{never} occur. \comm{Similarly, real biological systems are unlikely to satisfy the assumption of \emph{exactly} equal synthesis rate functions. In Appendix Sect.~\ref{sect: OL appendix} we show that offsets to the assumed synthesis rates preserve the conclusion that fluctuations diverge as $E\to 100\%$, as long as the offsets do not encode additional feedback.} %

Here we have included simultaneous de novo subunit synthesis to reconcile the observation of finite fluctuations despite 100\% assembly efficiency~\cite{Laurenti2018} with the prediction that fluctuations must diverge in this limit~\cite{Hilfinger2016a}. However, even in biological systems in which the synthesis of subunits is tightly coupled, subunit synthesis typically does not occur in a single chemical reaction. For example, transcriptional regulation is shared across mRNA transcripts expressed from operons encoding the subunits of protein complexes. However, this does not \comm{necessarily} lead to simultaneous protein synthesis events, as ribosomes \comm{typically} synthesize proteins individually~\cite{Green1997}. \comm{Similarly}, %
for bicistronic mRNA in eukaryotes, translational initiation occurs independently for each cistron~\cite{Lastra2005}. Another case of tightly coupled synthesis is the expression of microRNA-mRNA pairs from bicistronic DNA~\cite{Bartel2004,Posadas2014}. However, since RNA polymerases transcribe DNA to synthesize RNA molecules sequentially, RNA molecules will not enter the system at the exact same time even in such processes. For both microRNA-mRNA complexes and post-translationally assembled protein complexes it is thus likely that simultaneous ``production'' is primarily due to complex dissociation and that such systems are therefore unlikely to operate near $F = 100\%$. \comm{In contrast, some protein complexes in bacteria and eukaryotes seem to form through co-translated nascent protein chains that associate during elongation~\cite{Shieh2015,Kamenova2019,Bertolini2021, Badonyi2022}.
Indeed, such `co-co-translational' systems may achieve large $F$ values for the protein subunits if nascent chains rarely get translated independently.}

\comm{In addition to quantifying how simultaneous production affects the minimum subunit fluctuations,} \cref{eq: beta asymmetry} also quantifies how asymmetric degradation rates decrease the severity of the previously reported bound. However, regardless of the exact value of $\beta_2/\beta_1$ in a particular system, a trade-off exists between efficiency and fluctuations, with a lower bound on subunit fluctuations
that diverges as the assembly efficiency approaches 100\%.
While the bound vanishes for infinitely asymmetric degradation rates, energetic constraints for turnover rates limit this asymmetry in biological systems.
For example, in \textit{E.~coli}, mRNA half-lives \cite{Bernstein2002} suggest a realistic maximum $\beta_2/\beta_1$ of around $15$, for which we still see a significant trade-off, see Fig.~\ref{fig: single-input}(d).
Data obtained for arrested HeLa cells suggests a greater diversity of protein turnover rates across the proteome compared to mRNA turnover in \textit{E.~coli}~\cite{Cambridge2011}. However, relatively small variations were observed when analyzing turnover rates of subunits forming specific protein complexes, which suggests limited degradation rate asymmetries for protein subunits \comm{within a given complex}~\cite{Cambridge2011}. \comm{Nevertheless, we address the singular case of $\beta_2/\beta_1 = \infty$, $E = 1$ in Appendix \ref{sect: beta2/beta1 = infty, E=1}, showing that such a system is unstable and cannot reach a stationary state with finite variances.}

Since we are motivated to characterize systems without knowing their rate constants, \cref{eq: beta asymmetry} constrains systems for arbitrary values of $F$ and $E$, which are treated as system observables in our analysis. We do not consider how the fluctuations of specific systems change when specific rate constants change while others are held fixed. Previous work has performed such analyses for sRNA-RNA assembly processes~\cite{Mehta2008} and protein complex assembly processes~\cite{Hakkinen2013,Lord2019}. %
Note also, the fluctuation bound presented in \cref{eq: beta asymmetry} becomes loose 
for systems with strongly asymmetric degradation rates. Actual systems with $\beta_2>\beta_1$ have to exhibit even larger fluctuations due to a more severe bound, see Appendix~\ref{sect: beta appendix}.

\subsection{Systems with asymmetric complex stoichiometry}\label{sect: hoc}
So far we considered assembly processes in which the complex consists of exactly two subunits. However, protein complexes often contain multiple copies of individual subunits. For example, some ribosomal complexes are formed from subunits with 1:4 stoichiometry~\cite{Ma1981,Li2014}.
Studies in bacteria suggest that in such cases, 
complex-forming proteins are synthesized at rates that are proportional to their stoichiometry in the complex~\cite{Li2014}. 

\begin{figure}[ht]
    \centering
    \includegraphics[width = \linewidth]{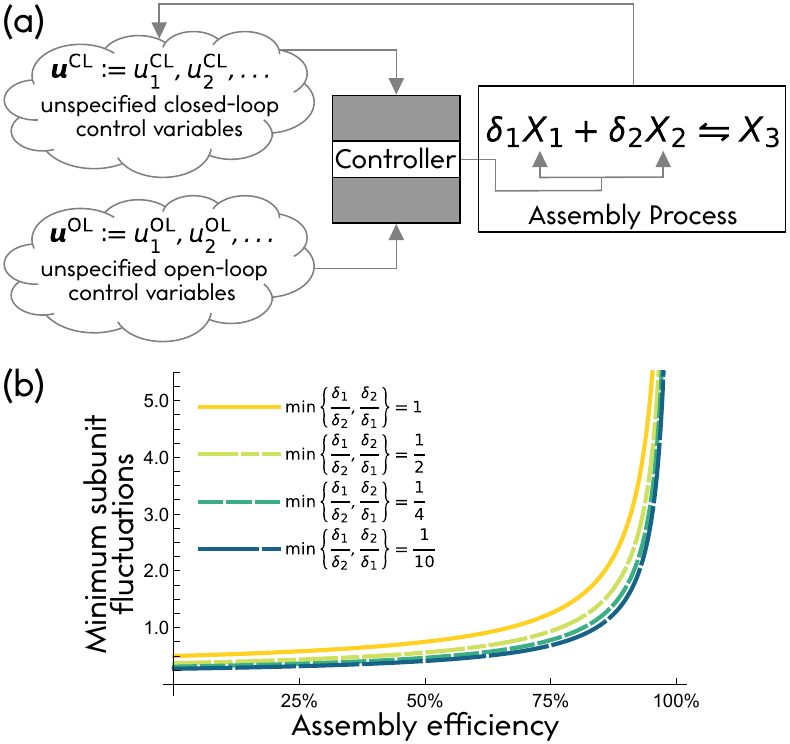}
    \caption{\textit{Asymmetric complex stoichiometry preserves the diverging efficiency-fluctuation trade-off for assembly processes under single-input control.} (a) We consider all possible assembly processes where two subunits $X_1,X_2$ form a complex $X_3$ with stoichiometry $\delta_1:\delta_2$ and the control of subunit synthesis rates is shared such that subunit synthesis fluxes are proportional to the complex stoichiometry, see \cref{eq: higher-order complexes}. (b) The minimum subunit fluctuations increase monotonically as a function of assembly efficiency $E$, see \cref{eq: hoc bound}. Plotted is the lower bound for different stoichiometric ratios while keeping $\beta_1 = \beta_2$ and $F = 0$. Subunit fluctuations are defined relative to the Poisson fluctuations subunits would exhibit with constant synthesis rates in the absence of complex formation. For all complexing stoichiometries subunit fluctuations diverge as the assembly efficiency approaches $100\%$. %
    }
    \label{fig: delta}
\end{figure}

We thus next consider a class of assembly processes where the complex consists of $\delta_1$ copies of $X_1$, $\delta_2$ copies of $X_2$, and subunits are individually synthesized with rates in proportion to this stoichiometry: 
\begin{equation}\label{eq: higher-order complexes}
    \begin{gathered}
        x_1  \overset{\delta_1 r_I}{\xrightarrow{\hspace*{1.1cm}}} x_1 + 1 \hspace{0.47cm} x_2  \overset{\delta_2 r_I}{\xrightarrow{\hspace*{1.1cm}}} x_2 + 1\vspace{-.3em}\\ \text{\emph{\footnotesize{individual subunit synthesis}}}\\
        x_1 \overset{\beta_1 x_1}{\xrightarrow{\hspace*{1.1cm}}} x_1-1 \hspace{0.47cm} x_2 \overset{\beta_2 x_2}{\xrightarrow{\hspace*{1.1cm}}} x_2-1\vspace{-.3em}\\ \text{\emph{\footnotesize{subunit degradation}}}\\
        (x_1, x_2 ,x_3) \overset{r_A}{\xrightarrow{\hspace*{1.1cm}}} (x_1 - \delta_1, x_2 - \delta_2, x_3 + 1)\vspace{-.3em}\\ \text{\emph{\footnotesize{complex assembly}}}\\
        (x_1, x_2 ,x_3) \overset{r_D}{\xrightarrow{\hspace*{1.1cm}}} (x_1 + \delta_1, x_2 + \delta_2, x_3 - 1)\vspace{-.3em}\\ \text{\emph{\footnotesize{complex dissociation}}}\\
        x_3 \overset{\beta_3 x_3}{\xrightarrow{\hspace*{1.1cm}}} x_3-1 \vspace{-.3em}\\ \text{\emph{\footnotesize{complex degradation}}}
    \end{gathered}
\end{equation}
where, as before, the rates
\begin{eqnarray*}
    r_I &= r_I(\boldsymbol{u}^\mathrm{OL},\boldsymbol{u}^\mathrm{CL}, \boldsymbol{x})\\
    r_A  &= r_A(\boldsymbol{u}^\mathrm{OL},\boldsymbol{u}^\mathrm{CL}, \boldsymbol{x})\\
    r_D &=r_D(\boldsymbol{u}^\mathrm{OL},\boldsymbol{u}^\mathrm{CL}, \boldsymbol{x})
\end{eqnarray*}
are unspecified such that they can depend in arbitrary ways on all components in the system and thus allow for arbitrary feedback, extrinsic noise, and open-loop control, see Fig.~\ref{fig: delta}(a). 

Assuming again $\beta_2 \geq \beta_1$ without loss of generality, and following the same approach as above (see Appendix~\ref{sect: delta generalization}), we find that the subunit fluctuations in this class of systems are bounded by
\begin{equation}\label{eq: hoc bound}
    \frac{1}{2}\left(\frac{\CV_1^2}{1/\langle x_1 \rangle} +  \frac{\CV_2^2}{1/\langle x_2 \rangle}\right)
    \geq \frac{f\left(\mfrac{\delta_1}{\delta_2}\right)}{3+\mfrac{\beta_2}{\beta_1}}\left(1+\frac{1-F}{1-E}\right)\ ,
\end{equation}
where
\begin{equation}\label{eq: f(delta ratio)}
    f\left(\frac{\delta_1}{\delta_2}\right) \coloneqq \frac{1 + \min\left\{\mfrac{\delta_1}{\delta_2},\mfrac{\delta_2}{\delta_1}\right\}}{2}\ .
\end{equation}
Thus, similar to increasing the degradation rate asymmetry, increasingly asymmetric stoichiometric ratios can reduce the severity of the bound. However, for any $F<100\%$, subunit fluctuations are constrained by a lower bound that diverges as the assembly efficiency approaches 100\%, see Fig.~\ref{fig: delta}(b). 

Note, technically the equations specifying $E$ and $F$ differ here compared to \cref{eq: E defn} and \cref{eq: F defn} due to the differences in step-sizes and rate functions. In particular, here $F$ quantifies the relative importance of complex dissociation to overall subunit synthesis since we have excluded simultaneous de novo synthesis reactions in the class of systems defined by \cref{eq: higher-order complexes}, such that subunit synthesis is under single-input control, see Appendix \crefrange{eq: phi_i appendix}{eq: F_i appendix}. Nevertheless, the interpretation of $E$ and $F$ remains as the assembly efficiency and simultaneous production fraction, respectively, see Appendix \crefrange{eq: E_i appendix}{eq: F_i appendix}. 

While experimental evidence suggests operonic protein synthesis fluxes are proportional to complex stoichiometry~\cite{Li2014}, actual biological processes might not exhibit the \emph{perfectly} proportional fluxes we assumed to define the class of systems in \cref{eq: higher-order complexes}. \comm{However, as for systems obeying ~\cref{eq: beta asymmetry}, we show in Appendix Sect.~\ref{sect: OL appendix} that offsets to the synthesis rates here preserve the conclusion that fluctuations diverge as $E\to 100\%$, as long as the offsets do not encode additional feedback. In contrast, in Section~\ref{sect: dual input cl results} we describe specific systems that avoid diverging subunit fluctuations in the high efficiency limit due to significantly different feedback regulation of subunit synthesis. However, further analysis will be required to determine whether \emph{slight} deviations from the assumed synthesis rate symmetries underlying the derivations of Eqs.~(\ref{eq: beta asymmetry},~\ref{eq: hoc bound}) could affect these fluctuation bounds.} 

\subsection{\comm{Systems with distinct subunit synthesis rates under open-loop control}}\label{sect: OL}
\begin{figure}[ht]
    \centering
    \includegraphics[width = 0.95\linewidth]{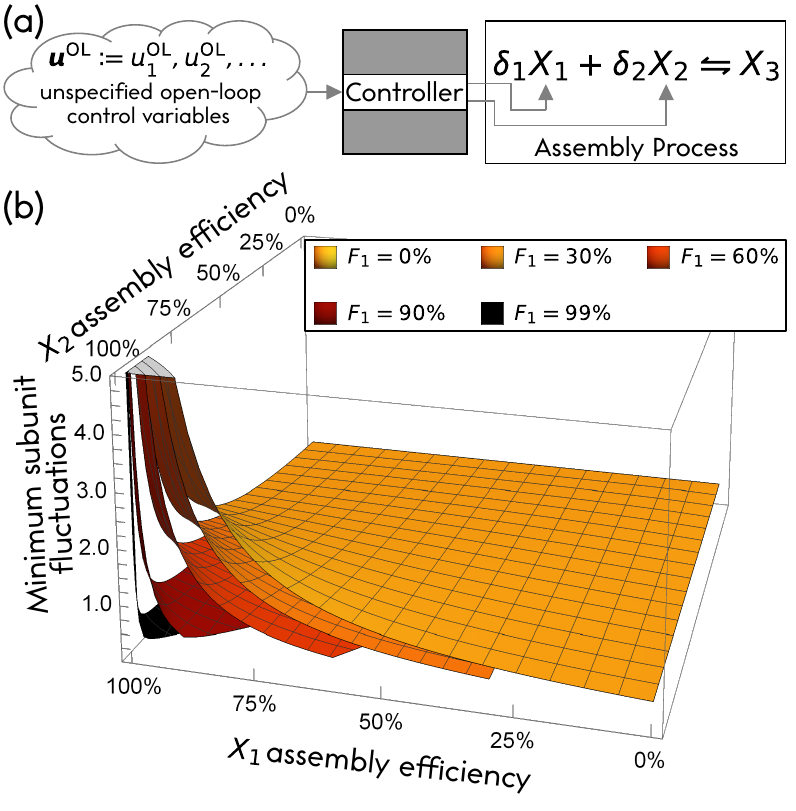}
    \caption{\textit{Assembly processes under dual-input open-loop control obey a diverging efficiency-fluctuation trade-off.} (a) We consider all possible assembly processes where two subunits $X_1,X_2$ form a complex $X_3$ with arbitrary stoichiometry and are subject to independent open-loop control. (b) For sufficiently large efficiencies, minimum subunit fluctuations increase monotonically with respect to the individual subunit efficiencies $E_1, E_2$. Plotted is the lower bound of \cref{eq: dual input OL bound} for different values of the simultaneous $X_1$ production ratio $F_1$ while keeping subunit degradation symmetric ($\beta_1 = \beta_2$). Subunit fluctuations are quantified relative to the Poisson fluctuations that subunits would exhibit with constant synthesis rates in the absence of complex formation. %
    For all values of $F_1 < 100\%$ subunit fluctuations diverge as $E_1,E_2 \to 100\%$. The depicted example bounds correspond to systems without simultaneous de novo subunit synthesis, i.e, with $r_S = 0$, such that $F_1$ quantifies the fraction of $X_1$ subunits entering the system through complex dissociation. In this limit the requirement that $F_2 \leq 1$ implies a restricted domain satisfying $E_2/E_1 \leq 1/F_1$, see Appendix \crefrange{eq: E_i appendix}{eq: F_i appendix}.
    }
    \label{fig: dual-input OL}
\end{figure}

So far we assumed that subunits are synthesized with rates that are proportional to their stoichometry in the complex. Next, we consider assembly processes in which the subunit synthesis rates are different but only depend on an unspecified set of upstream components that are not causally affected by $X_1,X_2,$ or $X_3$, i.e., the shared individual synthesis rate $r_I$ for $X_1$ and $X_2$ in \cref{eq: 2016 with co-synthesis - subunit dynamics} is replaced with
\begin{eqnarray*}
    r_1 &= r_1(\boldsymbol{u}^\mathrm{OL}) \\
    r_2 &= r_2(\boldsymbol{u}^\mathrm{OL}) \ ,
\end{eqnarray*}
respectively. Similarly, the reaction rate corresponding to simultaneous de novo synthesis is also assumed to not be influenced through a feedback loop, i.e., we have
\begin{eqnarray*}
    r_S &= r_S(\boldsymbol{u}^\mathrm{OL}) \ . 
\end{eqnarray*}
In contrast, the association and dissociation rates $r_A,r_D$ are left completely unspecified and allow for arbitrary feedback as before. Here we also allow for arbitrary complex stoichiometry where $\delta_1$ copies of $X_1$ and $\delta_2$ copies of $X_2$ are sequestered to the complex, as in \cref{eq: higher-order complexes}. This defines the class of all possible such assembly processes under dual-input open-loop control of subunit synthesis rates as illustrated in Fig.~\ref{fig: dual-input OL}(a). %

For this class of systems, the fraction of subunit molecules that form a complex at least once is not necessarily equal across $X_1,X_2$ since the subunit synthesis fluxes are not restricted to be in proportion to the stoichiometry of the complex. Thus, we define the following assembly efficiency parameters for $X_1$ and $X_2$, respectively:
\begin{equation}
    \begin{aligned}
        E_1 &\coloneqq \frac{\delta_1\langle r_A \rangle}{\delta_1\langle r_A \rangle + \beta_1 \langle x_1 \rangle}\\
        E_2 &\coloneqq \frac{\delta_2\langle r_A \rangle}{\delta_2\langle r_A \rangle + \beta_2 \langle x_2 \rangle}\ .
    \end{aligned}
\end{equation}
Similarly, the fraction of $X_1$ molecules that enter the system with an $X_2$ molecule may differ from the fraction of $X_2$ molecules that enter the system with an $X_1$ molecule. We thus specify the following simultaneous ``production'' fractions for $X_1$ and $X_2$, respectively: 
\begin{equation}
    \begin{aligned}
        F_1 &\coloneqq \frac{\langle r_S \rangle + \delta_1 \langle r_D \rangle}{\langle r_1 \rangle + \langle r_S \rangle + \delta_1\langle r_D \rangle}  \\
        F_2 &\coloneqq \frac{\langle r_S \rangle + \delta_2 \langle r_D \rangle}{\langle r_2 \rangle + \langle r_S \rangle + \delta_2\langle r_D \rangle}\ .
    \end{aligned}
\end{equation}

Again taking $\beta_2 \geq \beta_1$ without loss of generality, we obtain the following bound on subunit fluctuations:
\begin{multline}\label{eq: dual input OL bound}
     \frac{1}{2}\left(\frac{\CV_1^2}{1/\langle x_1 \rangle} + \frac{\CV_2^2}{1/\langle x_2 \rangle}\right) \geq \frac{\left(1 + \min\left\{\frac{\delta_1}{\delta_2},\frac{\delta_2}{\delta_1}\right\}\right)}{4\left(\frac{1}{\delta_1} + \frac{1}{\delta_2}\right)}\\
     \times \frac{\left[E_2(2 - E_1 - F_1)/\delta_2\right]+ \left[E_1(2 - E_2 - F_2)/\delta_1\right]}{E_1(1-E_2) + \left(\frac{\beta_2}{\beta_1} + 1 - \sqrt{\frac{\beta_2}{\beta_1}}\right)E_2(1-E_1)}\ ,
\end{multline}
see Appendix~\ref{sect: OL appendix}.

While the bound of \cref{eq: dual input OL bound} is algebraically more complex due to the asymmetry in subunit synthesis, it tells the same story as \cref{eq: beta asymmetry} and \cref{eq: hoc bound}: The right side of this inequality diverges for $E_1,E_2 \to 100\%$ as long as $F_1,F_2$ are not both exactly $100\%$, i.e., even in the presence of infinitesimally rare individual synthesis events subunit fluctuations must diverge in the 100\% efficiency limit. Increasing the fraction of subunits that enter the system simultaneously through increasing $F_1$ or $F_2$ reduces the lower bound on fluctuations for given efficiency levels $E_1,E_2$, e.g., see Fig.~\ref{fig: dual-input OL}(b). Additionally, similar to before, increasing the degradation rate asymmetry $\beta_2/\beta_1$ reduces the lower bound for fixed values of $E_1,E_2,F_1,F_2$. Likewise, increasingly asymmetric complex stoichiometry encoded by the parameters $\delta_1,\delta_2$ reduces the lower bound, as before. 

\comm{While our main results Eqs.~(\ref{eq: beta asymmetry},~\ref{eq: hoc bound},~\ref{eq: dual input OL bound}) theoretically constrain broad classes of complex-forming subunits in cellular processes,  experimentally testing these bounds remains an open problem. While the overall abundance of intracellular components is experimentally quantifiable (e.g., by combining fluorescent probes with high-resolution microscopy~\cite{Longo2006}, flow cytometry~\cite{Porter2017}, or single-molecule approaches~\cite{Raj2009}), direct estimation of the assembly efficiency would require tracking the fate of individual subunits. %
This presents a challenge to directly estimating assembly efficiencies. However, these quantities may be estimated indirectly from stationary abundance distributions \emph{if} biochemical parameters relating to the assembly process are well-characterized -- specifically, the kinetics of assembly and dissociation as well as degradation kinetics for the complex and subunits (but not the subunit synthesis rates). 
It would be interesting to perform such an indirect estimation of assembly efficiencies for systems where noisy subunit dynamics have been shown to underlie cellular decision making, such as the switching between biofilm-forming and motile states of \emph{B.~subtilis}~\cite{Norm2013, Lord2019}.}

\subsection{\comm{Systems with distinct subunit synthesis rates and closed-loop control}}\label{sect: dual input cl results}
The results of Eqs.~(\ref{eq: beta asymmetry},~\ref{eq: hoc bound},~\ref{eq: dual input OL bound}) establish that only systems with unequal subunit synthesis rates \textit{and} feedback control can possibly avoid infinite subunit fluctuations in assembly processes as assembly efficiency approaches 100\%. \comm{A potential biological example of such a system is sequestration-based inhibition of transcriptional activators in circadian oscillators~%
\cite{Kim2016}, where it is desirable to limit noise to ensure robust oscillation periods~\cite{Barkai2000}.}

Next, we analyze which properties of such systems are sufficient to achieve finite subunit fluctuations in the 100\% efficiency limit. Even though we now consider fully specified example systems, the stationary fluctuations cannot be solved analytically due to the nonlinearity of the system. We thus, use exact numerical simulations~\cite{doob1945,Gillespie1977} to demonstrate the ability of these systems to exhibit finite fluctuations in the limit of 100\% assembly efficiency. \comm{A handful of simulations for specific parameter sets cannot possibly characterize the behavior of all possible systems incorporating dual-input control of production rates and feedback. However, the following numerical simulations establish a constructive proof of existence of controllers that permit finite stationary state fluctuations  at $E=100\%$, which is impossible for systems constrained by the bounds of Eqs.~(\ref{eq: beta asymmetry},~\ref{eq: hoc bound},~\ref{eq: dual input OL bound}).} %

In each example we consider, the reactions involving the complex $X_3$ are given by mass-action kinetics
\begin{equation}\label{eq: assembly dynamics}
    \begin{gathered}
        (x_1,\ x_2,\ x_3) \overset{\gamma x_1 x_2}{\xrightarrow{\hspace*{1.1cm}}} (x_1 - 1, \ x_2 - 1,\ x_3 + 1) \vspace{-.3em}\\ \text{\emph{\footnotesize{complex assembly}}}\\
        (x_1,\ x_2,\ x_3) \overset{\lambda x_3}{\xrightarrow{\hspace*{1.1cm}}} (x_1 + 1, \ x_2 + 1,\ x_3 - 1) \vspace{-.3em}\\ \text{\emph{\footnotesize{complex dissociation}}}\\
        x_3 \overset{\beta_3 x_3}{\xrightarrow{\hspace*{1.1cm}}} x_3 - 1\ . \,\vspace{-.3em}\\ \text{\emph{\footnotesize{complex degradation}}}
    \end{gathered}
\end{equation}
Each example system is then defined by specifying a control strategy which regulates subunit levels through two distinct synthesis rates $r_1,r_2$, where at least one encodes feedback control. 

\begin{figure}[ht]
    \centering
    \includegraphics[width = \textwidth]{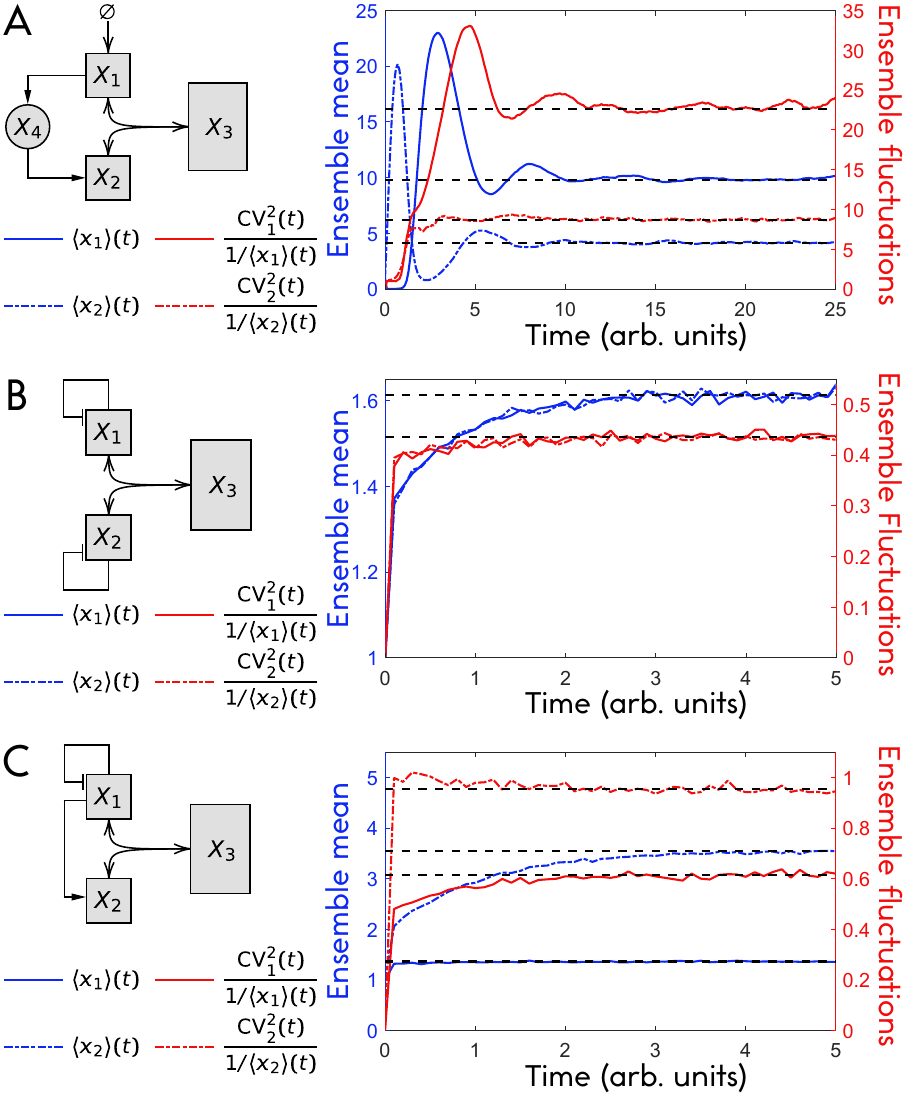}
    \caption{\textit{%
    Systems with distinct subunit synthesis rates and feedback can avoid diverging fluctuations in the 100\% efficiency limit
    }   
    Numerical simulation data for three examples of feedback control and distinct subunit synthesis rates. Right column shows time-dependent ensemble statistics across 8000 independent simulations with 100\% assembly efficiency, as detailed in Appendix~\ref{sect: numerics}. 
    Each example shows convergence of the mean subunit abundance (left axis, blue) to the numerically determined long-time average of an example trace (black dotted lines). Similarly, we observe convergence of the subunit fluctuations to a finite long-time limit relative to Poisson fluctuations (right axis, red).
    (a) Antithetic integral feedback example given by \crefrange{eq: assembly dynamics}{eq: AIF module-2}. (b) Symmetric direct feedback example given by Eqs~(\ref{eq: assembly dynamics},~\ref{eq: symmetric rates}). (c) Asymmetric direct feedback example given by Eqs~(\ref{eq: assembly dynamics},~\ref{eq: asymmetric rates}). See Appendix~\ref{sect: numerics} for parameter values used in the numerical simulations.}
    \label{fig: dual-input CL}
\end{figure}

First, we consider a simple realization of antithetic integral feedback~\cite{Briat2016} given by \cref{eq: assembly dynamics} along with the following subunit synthesis dynamics:
\begin{equation}\label{eq: AIF module-1}
    \begin{gathered}
        x_1 \overset{\mu}{\xrightarrow{\hspace*{1.1cm}}} x_1+1 \hspace{18pt} x_2 \overset{\theta x_4}{\xrightarrow{\hspace*{1.1cm}}} x_2+1\ ,
    \end{gathered}
\end{equation}
where $X_4$ is a closed-loop control variable which senses the levels of $X_1$ through the following dynamics: 
\begin{equation}\label{eq: AIF module-2}
    \begin{gathered}
        x_4 \overset{kx_1}{\xrightarrow{\hspace*{1.1cm}}} x_4+1 \hspace{18pt} x_4\overset{\beta_4 x_4}{\xrightarrow{\hspace*{1.1cm}}} x_4-1\,
    \end{gathered}
\end{equation}
see Fig.~\ref{fig: dual-input CL}(a). Furthermore, we set the subunit degradation rates $\beta_1,\beta_2$ equal to zero which ensures that the assembly efficiency is at 100\%.

Antithetic integral feedback has received much attention (e.g.~\cite{Briat2016, Briat2018,Qian2018,Olsman2019a,Aoki2019,Frei2022,Filo23}) due to the remarkable property that, in the 100\% efficiency limit, it achieves robust perfect adaptation for the stationary average of a component of interest~\cite{Briat2016, Aoki2019}. This property holds so long as the overall system is stable and has been demonstrated in engineered living systems~\cite{Aoki2019,Frei2022}. Here, this property is conferred to the component $X_4$, such that 
$\langle x_4 \rangle = \mu/\theta$ %
no matter the value of other parameters~\cite{Briat2016,Aoki2019}.

However, our primary interest here is not the adaptation of average abundances controlled by the antithetic integral feedback module, but rather how the module exhibits finite subunit fluctuations despite operating at 100\% assembly efficiency. Previous mathematical analysis has shown that ergodic wide-sense stationary states exist for certain realizations of antithetic integral feedback in the absence of complex dissociation~\cite{Briat2016}. Here we numerically demonstrate the system defined by \crefrange{eq: assembly dynamics}{eq: AIF module-2} can maintain this property even in the presence of inevitable complex dissociation, see Fig.~\ref{fig: dual-input CL}(a). 

To test whether antithetic integral feedback is a unique control strategy that allows for 
finite
subunit fluctuations at 100\% assembly efficiency, we next consider two additional examples of control strategies combining feedback with distinct subunit synthesis rates. First, we consider a fully symmetric case where each subunit represses its own synthesis through the following dynamics: 
\begin{equation}\label{eq: symmetric rates}
    \begin{gathered}
        x_1 \overset{f(x_1)}{\xrightarrow{\hspace*{1.1cm}}} x_1+1 \hspace{18pt} x_2 \overset{f(x_2)}{\xrightarrow{\hspace*{1.1cm}}} x_2+1
    \end{gathered}
\end{equation}
where $f(x) = \nu x^n/(x^n + \kappa^n),\ n<0$ is a repressing Hill function~\cite{Weiss1997}. We numerically establish that this control scheme can also lead to ergodic wide-sense stationary states for the system composed of Eqs.~(\ref{eq: assembly dynamics},~\ref{eq: symmetric rates}), and avoids diverging subunit fluctuations despite 100\% efficient assembly, see Fig.~\ref{fig: dual-input CL}(b). %

Finally, we consider a case where the subunit $X_1$ represses its own synthesis and activates the synthesis of the other subunit $X_2$ through the following dynamics: 
\begin{equation}\label{eq: asymmetric rates}
    \begin{gathered}
        x_1 \overset{f_1(x_1)}{\xrightarrow{\hspace*{1.1cm}}} x_1+1 \hspace{18pt} x_2 \overset{f_2(x_1)}{\xrightarrow{\hspace*{1.1cm}}} x_2+1\vspace{-.3em}
    \end{gathered}
\end{equation}
where $f_1(x) = \nu_1 x^{n_1}/(x^{n_1} + \kappa^{n_1}), \ n_1 <0$ is a repressing Hill function, while $f_2(x) = \nu_2x^{n_2}/(x^{n_2} + \kappa^{n_2}), \ n_2 > 0$ is an activating Hill function~\cite{Weiss1997}. Since a surplus of $X_2$ leads to increased sequestration of $X_1$ through the assembly reaction, this configuration leads to $X_1$-mediated negative feedback on both subunits. As in the previous two examples, we numerically demonstrate that this control scheme can achieve ergodic wide-sense stationary states for the system composed of Eqs.~(\ref{eq: assembly dynamics},~\ref{eq: asymmetric rates}), thus avoiding diverging subunit fluctuations despite 100\% efficient assembly, see Fig.~\ref{fig: dual-input CL}(c). 

The above examples of control strategies achieve finite fluctuations though diverse mechanisms: through an intermediate species affecting only one subunit, through direct symmetric self-feedback on each subunit, and through direct asymmetric feedback where a single subunit regulates itself and the other subunit. The numerical results for these systems demonstrate that antithetic integral feedback is not unique in its ability to avoid diverging subunit fluctuations in the limit of 100\% assembly efficiency. We further verified this property is not in general singular -- i.e., that fluctuations strengths near 100\% efficiency can be similar to those at exactly 100\% efficiency. \comm{Specifically, we simulated the fully symmetric model (Eqs.~(\ref{eq: assembly dynamics},~\ref{eq: symmetric rates})) with increasingly small subunit degradation rates to approach $E_1=E_2=100\%$ in two ways; i) as $E_1=E_2$ approaches $1$ and ii) as $E_1$ approaches $1$ with $E_2=1$ (which is equivalent to $E_2$ approaching $1$ with $E_1 = 1$ in this system by symmetry). In each case the fluctuation strengths change marginally across the observed efficiency values. For example, the numerical fluctuation strengths increased by $0.25\%$ for a $0.6\%$ decrease in $E_1=E_2$ from $100\%$ and become numerically indistinguishable from the $E_1=E_2 =100\%$ fluctuations for $E_1 = E_2 \approx 99.99\%$. These numerical data support our expectation that stable systems under dual input control with feedback will exhibit similar fluctuations near perfect efficiency as at perfect efficiency.}

\section{Discussion \& conclusions}
Complex electronic and mechanical systems can be understood and designed in terms of well-characterized properties of individual modules. Such a modular understanding of cells would \comm{clarify the constraints on natural biological control} and greatly improve our ability to engineer cellular processes~\cite{DelVecchio2018,GRUNBERG202041}. However, %
complex biochemical networks can exhibit emergent properties that are not predicted by analyzing their components in isolation~\cite{Cardinale2012}. 

Here, we utilize general statistical invariants for pairs of stochastic variables to constrain the behavior of a particular reaction motif. %
This approach permits arbitrary interactions \comm{between the specified motif and} %
\comm{a larger unspecified} network~\cite{Hilfinger2016a}. %
However, for such \comm{constraints} to establish biological principles, we need to show that systems do not exhibit drastically different behaviour if they deviate slightly from the precise mathematical \comm{definitions of the specified reaction motifs.} %

By making changes to the specified reactions in the previously discussed class of assembly processes~\cite{Hilfinger2016a} we here report such a ``sensitivity to assumptions'' analysis. %
\comm{Our analysis shows} that much broader classes of assembly processes \comm{than previously considered} are limited by an efficiency-fluctuation trade-off in which fluctuations diverge in the limit of 100\% assembly efficiency. 

\comm{In particular, we consider, distinct regulation of subunit production rates where at least one encodes feedback is required to avoid an efficiency-fluctuation trade-off which diverges in the 100\% efficiency limit. The class of antithetic integral feedback systems is a previously known example of such control~\cite{Briat2016, Aoki2019}. We provide two additional numerical examples of systems which maintain finite fluctuations as assembly efficiency approaches 100\%. In these examples, both subunit synthesis rates are feedback controlled with significantly different control functions for each subunit. The numerical examples presented illustrate that there are several ways to avoid the diverging fluctuations as efficiency approaches 100\%. We also show, however, that adding open-loop control offsets to the assumed shared control underlying the derivations of Eqs.~(\ref{eq: beta asymmetry},~\ref{eq: hoc bound}) cannot avoid diverging fluctuations in the 100\% efficiency limit. While this implies that not all deviations from the assumptions of shared control avoid an efficiency-fluctuation trade-off, it remains an open question as to how these fluctuation bounds change for small deviations from the assumption of shared control.
Future work is required to analyze whether minute deviations from shared subunit control could maintain finite fluctuations as efficiency approaches 100\%.}

Our definitions of generalized assembly motifs %
include several \comm{other} idealized assumptions which biological systems may deviate from. In particular, our analysis strictly only applies to control schemes that directly regulate the synthesis rates of complex-forming subunits. This is realistic to describe assembly processes between RNA species such as microRNA-RNA complexes~\cite{Bartel2004,Posadas2014}. However, protein synthesis is often regulated at the level of transcription, where our models correspond to infinitely fast mRNA-dynamics such that control signals are not delayed or washed out by stochastic fluctuations in mRNA-levels. 

In the case of complexes with multiple subunit copies, we assume association and dissociation are well-described in an ``infinite cooperativity'' limit such that all copies of each subunit are sequestered to or released from the complex simultaneously. However, %
\comm{self-assembly of many components is more realistically modeled by considering sequential association and dissociation events~\cite{Gartner2020, Gartner2022, Jhaveri2024}}.

\comm{Furthermore, our analysis is restricted to systems with only two distinct subunit types, excluding a wide range of important macromolecular complexes such as the cellular translation machinery.}
Our analysis also assumes strictly uncontrolled, first-order degradation rates. \comm{Considering the more general case of controlled degradation rates for complex-forming subunits will be important for understanding fluctuations and control in a range of biological systems, including toxin-antitoxin systems where catalytic degradation of subunits is actively regulated~\cite{Page2016} or sigma-antisigma modules in which antisigma factors undergo sequestration reactions with both sigma factors and anti-antisigma factors~\cite{Hughes1998}.} 

Future work %
\comm{should thus} analyze the effect of \comm{i)} transcriptional regulation with finite mRNA timescales on protein subunit dynamics, \comm{ii)} sequential association/dissociation processes for complexes with \comm{more than two} subunit types, and \comm{iii)} non-linear subunit degradation rates and degradation rate control. Furthermore, future work should incorporate cell cycle dynamics and spatial heterogeneity to better understand how well our suggested general trade-off constrains fluctuations in actual biological systems.

\section*{Acknowledgements}
We thank Raymond Fan, Seshu Iyengar, Linan Shi, and Euan Joly-Smith for many helpful discussions and suggestions. This work was supported by the Natural Sciences and Engineering Research Council of Canada and a New Researcher Award from the University of Toronto Connaught Fund. BK gratefully acknowledges funding from the University of Toronto Faculty of Arts \& Science Top Doctoral Fellowship. Computational facilities for this work were generously provided by Compute Ontario and the Digital Research Alliance of Canada.

\section*{Author Contributions}
BK - Conceptualization, Data curation, Formal analysis, Investigation, Methodology, Software, Visualization, Writing - original draft, Writing - review and editing. AH - Conceptualization, Methodology, Resources, Supervision, Writing - review and editing. 

\appendix

\section*{Appendix}
Here we present derivations of the mathematical results presented in the main text. Additionally, we describe the numerical simulation procedure used to generate data for the dual-input control systems with feedback. 

\section{Stochastic dynamics of incompletely specified chemical reaction networks}\label{sect: dynamics of incompletely specified systems}
\comm{
Because we consider incompletely specified biochemical reactions networks in which the components of interest $\boldsymbol{x}$ may depend on components $\boldsymbol{u}$ whose dynamics is not specified (see \cref{eq: jump process}) we cannot analyze the system's complete chemical master equation but analyze the following marginalized dynamics for the probability $P(\boldsymbol{x};t)$ distribution of components of interest~\cite{Hilfinger2016a}
\begin{multline}\label{eq: cCME}
    \frac{\mathrm{d}P(\boldsymbol{x};t)}{\mathrm{d}t} \\
    = \sum_{k=1}^m\left[\left\lb\mathcal{R}_k|\boldsymbol{x}-\boldsymbol{d}_k\right\rb P(\boldsymbol{x}-\boldsymbol{d}_k;t) - \left\lb\mathcal{R}_k | \boldsymbol{x}\right\rb P(\boldsymbol{x};t) \right]\ ,
\end{multline}
where for notational simplicity we suppress the functional dependence of the reaction rates $\mathcal{R}_k$ on the specified and unspecified components.}

\comm{The conditional rates are impossible to determine without specifying the dynamics of all unspecified components that directly or indirectly affect the components of interest. 
However, exact statistical moment invariants can be derived from \cref{eq: cCME} for incompletely specified systems at stationarity}. Such invariants are the key ingredient in deriving general bounds for classes of sparsely characterizes systems here and in previous work~\cite{Hilfinger2016a, Hilfinger2016, Euan2021, Euan2024}. For example, the time-evolution of the first-order ensemble moments can be obtained directly from the marginalized CME~\cite{Lestas2008,Hilfinger2016a}:%
\begin{equation}\label{eq: mean evolution}
    \frac{\mathrm{d}\langle x_i \rangle(t)}{\mathrm{d}t} = \left\langle R_i^+\right\rangle(t) - \left\langle R_i^-\right\rangle(t)\ ,
\end{equation}
where $R_i^+, R_i^-$ are the total birth and death fluxes, respectively, of $X_i$ molecules given the current state of the system $\boldsymbol{x}$ and $\langle \cdot \rangle$ denotes an ensemble average, i.e.,
\begin{equation}\label{eq: ensemble average}
    \langle f(\boldsymbol{x}) \rangle(t) \coloneqq \sum_{j=1}^n\sum_{x_j=0}^\infty f(\boldsymbol{x}) P(\boldsymbol{x};t)\ ,
\end{equation}
for any function $f$ of the state-vector $\boldsymbol{x}$. Mathematically, the birth and death fluxes are given as
\begin{equation}
\begin{aligned}
    \comm{R_i^+(\boldsymbol{x})} &\comm{\coloneqq \sum_{k\in\left\{1,2,\ldots,m\right\}:d_{ik}>0}d_{ik}\left \lb \mathcal{R}_k | \boldsymbol{x} \right\rb}\\ 
    \comm{R_i^-(\boldsymbol{x})} &\comm{\coloneqq \sum_{k\in\left\{1,2,\ldots,m\right\}:d_{ik}<0}|d_{ik}|\left \lb\mathcal{R}_k |  \boldsymbol{x} \right\rb} \ .
\end{aligned}
\end{equation}
Similarly, the time evolution the of the ensemble covariance between components $X_i$ and $X_j$ (or variance if $i=j$) can be obtained directly from the CME and must obey
\begin{multline}\label{eq: covariance evolution}
    \frac{\mathrm{d}\Cov(x_i,x_j)(t)}{\mathrm{d}t} = \left\langle\sum_{k=1}^md_{ik}d_{jk} \mathcal{R}_k\right\rangle(t) \\
    + \Cov(R_i^+ - R_i^-, y_j)(t) + \Cov(R_j^+ - R_j^-, y_i)(t)\ .
\end{multline}

In the next section we describe the statistical moment invariants constraining stationary systems by considering these moment evolution equations at stationarity. Note, that we suppress the functional dependence of the birth and death fluxes in the covariance equations for notational simplicity.  

\section{Statistical moment invariants for stationary stochastic processes}\label{sect: general relations}
We consider wide-sense stationary processes such that the time evolution of the first (\cref{eq: mean evolution}) and second-order ensemble moments (\cref{eq: covariance evolution}) are at stationarity. Many relevant processes in biology also possess the property of ergodicity which ensures the temporal statistics for a sample trajectory match the ensemble statistics of the process in the long-time limit -- i.e., for any function $f$ of the state-vector, ergodicity provides the following at stationarity:
\begin{equation}
    \left\langle f(\boldsymbol{x})\right\rangle = \left\langle f(\boldsymbol{x}) \right\rangle^*\ ,
\end{equation}
where $ \left\langle f(\boldsymbol{x})\right\rangle$ is the $t \to \infty$ limit of the ensemble average defined by \cref{eq: ensemble average} and
\begin{equation}\label{eq: long-time averaging}
    \left\langle f(\boldsymbol{x}) \right\rangle^* \coloneq \lim_{t\to \infty}\frac{1}{t} \int_{0}^t f(\boldsymbol{x}(s))ds
\end{equation}
is the $t\to \infty$ limit of the time-average of a sample trajectory. Assuming ergodicity, all of our results can be interpreted either at the level of cell-to-cell variability in a population of genetically-identical cells in a shared environment or at the level of temporal variability in a single-cell. 

\comm{Previous work~\cite{Hilfinger2016a} showed that the assumption of stationarity can be relaxed to extend the following moment invariants to time averages of time-dependent conditional ensemble averages of non-diverging systems whose probability distributions are time-dependent -- e.g., cyclo-stationary processes that exhibit oscillatory behavior due to the cell cycle or circadian clocks. For brevity, we focus on illustrating the technique for ergodic stationary states below. For such systems, \cref{eq: mean evolution} implies}
\begin{equation}\label{eq: flux balance}
    \lb R_i^+\rb = \lb R_i^-\rb \eqqcolon \lb R_i^\pm\rb\ .
\end{equation}
for all components $X_i$. 

Before considering the stationary covariance relation we first define additional quantities to characterize the dynamics. First, we define the average co-step-size $\langle s_{ij} \rangle$ as the average change in $X_j$-abundance as a $X_i$ molecule is made or degraded, where the change is negative if one is made while the other is degraded. Mathematically, this is given by
\begin{equation}\label{eq: co-step sizes}
    \langle s_{ij} \rangle = \sum_{k=1}^m \frac{d_{ik}\langle \mathcal{R}_k(\boldsymbol{x})\rangle}{2\langle R_i^\pm \rangle}d_{jk}
\end{equation}
For $i=j$, this is simply the average jump size in $X_i$-levels across all reactions that change $X_i$ levels. Additionally, we define $\tau_i$ as the average life-time of a $X_i$ molecule which is related to average abundances and fluxes through Little's law from queuing theory~\cite{Little, Little2011} through
\begin{equation}\label{eq: Little's Law}
\begin{aligned}
    \tau_i &=\frac{\langle x_i \rangle^*}{\langle R_i^\pm \rangle^*} \\
    &= \frac{\langle x_i \rangle}{\langle R_i^\pm \rangle}\ , 
\end{aligned}
\end{equation}
where the second equality follows from ergodicity. 

Combining \crefrange{eq: flux balance}{eq: Little's Law} with \cref{eq: covariance evolution} at stationarity gives the previously reported (co)variance balance relation~\cite{Hilfinger2016a}
\begin{equation}\label{eq: covariance balance}
    \boldsymbol{U} + \boldsymbol{U}^T = \boldsymbol{D}
\end{equation}
for the matrices with entries
\begin{equation}
\begin{aligned}
    U_{ij} &= \frac{1}{\tau_j}\frac{\Cov(x_i, R_j^- - R_j^+)}{\left\lb x_i \right\rb \left\lb R_j^\pm \right\rb}\\
    D_{ij} &= \frac{1}{\tau_i}\frac{\left\lb s_{ij} \right\rb}{\left\lb x_j \right\rb} + \frac{1}{\tau_j}\frac{\left\lb s_{ji} \right\rb}{\left\lb x_i \right\rb}\ ,
\end{aligned}
\end{equation}
for all pairs of components $X_i, X_j$ whose state updates with Markovian dynamics, regardless of arbitrary, potentially non-Markovian, variables affecting the transition rates, as specified by \cref{eq: jump process}~\cite{Hilfinger2016a}.

\section{Applying stationary invariants to generalized assembly processes}\label{sect: assembly invariants}
We now apply these relations to the following class of systems, which includes all classes of systems discussed in the main text:
\begin{equation}\label{eq: super class}
    \begin{gathered}
        x_1  \overset{r_1(\boldsymbol{u}^\mathrm{OL},\boldsymbol{u}^\mathrm{CL}, \boldsymbol{x})}{\xrightarrow{\hspace*{1.1cm}}} x_1 + 1 \hspace{0.47cm} x_2  \overset{r_2(\boldsymbol{u}^\mathrm{OL},\boldsymbol{u}^\mathrm{CL}, \boldsymbol{x})}{\xrightarrow{\hspace*{1.1cm}}} x_2 + 1\\
         (x_1,x_2) \overset{r_S(\boldsymbol{u}^\mathrm{OL},\boldsymbol{u}^\mathrm{CL}, \boldsymbol{x})}{\xrightarrow{\hspace*{1.1cm}}} (x_1+1,x_2+1)\\
        x_1 \overset{\beta_1 x_1}{\xrightarrow{\hspace*{1.1cm}}} x_1-1 \hspace{0.47cm} x_2 \overset{\beta_2 x_2}{\xrightarrow{\hspace*{1.1cm}}} x_2-1\\
        (x_1, x_2 ,x_3) \overset{r_A(\boldsymbol{u}^\mathrm{OL},\boldsymbol{u}^\mathrm{CL}, \boldsymbol{x})}{\xrightarrow{\hspace*{1.1cm}}} (x_1 - \delta_1, x_2 - \delta_2, x_3 + 1)\\
        (x_1, x_2 ,x_3)\overset{r_D(\boldsymbol{u}^\mathrm{OL},\boldsymbol{u}^\mathrm{CL}, \boldsymbol{x})}{\xrightarrow{\hspace*{1.1cm}}} (x_1 + \delta_1, x_2 + \delta_2, x_3 - 1) \\
        x_3 \overset{\beta_3 x_3}{\xrightarrow{\hspace*{1.1cm}}} x_3-1\ .
    \end{gathered}
\end{equation}
The flux balance equations (\cref{eq: flux balance}) for each of $i = 1,2$ gives the following for this class of systems:
\begin{equation}\label{eq: R_i}
    \begin{aligned}
        \langle R_i^\pm \rangle &= \langle r_i \rangle + \langle r_S \rangle + \delta_i \langle r_D \rangle \\ 
        &= \beta_i \langle x_i \rangle + \delta_i \langle r_A \rangle \ .
    \end{aligned}
\end{equation}
Using the bilinearity of covariance to expand individual terms, the covariance balance equations (\cref{eq: covariance balance}) for $(i,j) \in \left\{(1,1),(2,2),(1,2)\right\}$ gives the following for this class of systems:
\begin{multline}\label{eq: cb11}
    \frac{\langle s_{11} \rangle}{\langle x_1 \rangle} = (1-E_1)\CV_1^2 + E_1\frac{\Cov(r_A,x_1)}{\langle r_A \rangle \langle x_1 \rangle} \\- (1-F_1)\frac{\Cov(r_1,x_1)}{\langle r_1 \rangle \langle x_1 \rangle} \\ - \psi_1\frac{\Cov(r_D,x_1)}{\langle r_D \rangle \langle x_1 \rangle} - \phi_1\frac{\Cov(r_S,x_1)}{\langle r_S \rangle \langle x_1 \rangle} 
\end{multline}
\begin{multline}\label{eq: cb22}
    \frac{\langle s_{22} \rangle}{\langle x_2 \rangle} = (1-E_2)\CV_2^2 + E_2\frac{\Cov(r_A,x_2)}{\langle r_A \rangle \langle x_2 \rangle} \\- (1-F_2)\frac{\Cov(r_2,x_2)}{\langle r_2 \rangle \langle x_2 \rangle} \\ - \psi_2\frac{\Cov(r_D,x_2)}{\langle r_D \rangle \langle x_2 \rangle} - \phi_1\frac{\Cov(r_S,x_2)}{\langle r_S \rangle \langle x_2 \rangle} 
\end{multline}
\begin{multline}\label{eq: cb12}
    \frac{\langle s_{12} \rangle}{\tau_1\langle x_2 \rangle} + \frac{\langle s_{21} \rangle}{\tau_2\langle x_1 \rangle} = \left[\frac{1-E_1}{\tau_1} + \frac{1-E_2}{\tau_2}\right]\eta_{12} \\+ \frac{E_1}{\tau_1}\frac{\Cov(r_A,x_2)}{\langle r_A \rangle \langle x_2 \rangle} + \frac{E_2}{\tau_2}\frac{\Cov(r_A,x_1)}{\langle r_A \rangle \langle x_1 \rangle} \\ 
    - \frac{1-F_1}{\tau_1}\frac{\Cov(r_1,x_2)}{\langle r_1 \rangle \langle x_2 \rangle} - \frac{1-F_2}{\tau_2}\frac{\Cov(r_2,x_1)}{\langle r_2 \rangle \langle x_1 \rangle} \\ 
    - \frac{\psi_1}{\tau_1}\frac{\Cov(r_D,x_2)}{\langle r_D \rangle \langle x_2 \rangle} - \frac{\psi_2}{\tau_2}\frac{\Cov(r_D,x_1)}{\langle r_D \rangle \langle x_1 \rangle} \\
    - \frac{\phi_1}{\tau_1}\frac{\Cov(r_S,x_2)}{\langle r_S \rangle \langle x_2 \rangle} - \frac{\phi_2}{\tau_2}\frac{\Cov(r_S,x_1)}{\langle r_S \rangle \langle x_1 \rangle} \ ,
\end{multline}
where for each $i= 1,2$ we define
\begin{align}
        E_i &\coloneqq \frac{\delta_i\langle r_A \rangle}{\langle R_i^\pm \rangle}\label{eq: E_i appendix}\\
        \phi_i &\coloneqq \frac{\langle r_S \rangle}{\langle R_i^\pm \rangle}\label{eq: phi_i appendix}\\
        \psi_i &\coloneqq \frac{\delta_i\langle r_D \rangle}{\langle R_i^\pm \rangle}\label{eq: psi_i appendix}\\
        F_i &\coloneqq \phi_i + \psi_i\label{eq: F_i appendix} \ .
\end{align}
and
\begin{equation}
    \eta_{12} \coloneqq \frac{\Cov(x_1,x_2)}{\langle x_1 \rangle \langle x_2 \rangle}\ .
\end{equation}
Additionally, $\CV_i$ is the coefficient of variation for the component $X_i$, as defined in \cref{eq: CV}. 

While additional moment invariants can be obtained by considering pairs of components including $X_3$, we do not need these relations for the derivations presented here. 

\section{Derivation of a general bound on subunit fluctuations}\label{sect: general bound}
Here we derive a general bound on subunit fluctuations from the covariance balance relations \crefrange{eq: cb11}{eq: cb12}. To do so, we first rescale \cref{eq: cb11} and \cref{eq: cb22} by $E_2/(E_1\tau_2)$ and $E_1/(E_2\tau_1)$, respectively. Substracting the rescaled relations from \cref{eq: cb12} yields:
\begin{multline}\label{eq: gbd1}
     \frac{(1-E_1)E_2}{E_1\tau_2}\CV_1^2 + \frac{(1-E_2)E_1}{E_2\tau_1}\CV_2^2 = \Lambda \\
    + \frac{E_2\langle s_{11}\rangle  - E_1\langle s_{21} \rangle}{E_1\tau_2\langle x_1 \rangle} + \frac{E_1\langle s_{22}\rangle  - E_2\langle s_{12} \rangle}{E_2\tau_1\langle x_2 \rangle}  \\
    +\left[\frac{1-E_1}{\tau_1} + \frac{1-E_2}{\tau_2}\right]\eta_{12}\ ,
\end{multline}
where
\begin{multline}\label{eq: LAMBDA defn}
    \Lambda \coloneqq \frac{(1-F_1)E_2}{E_1\tau_2}\frac{\Cov(r_1,x_1)}{\langle r_1 \rangle\langle x_1\rangle} + \frac{(1-F_2)E_1}{E_2\tau_1}\frac{\Cov(r_2,x_2)}{\langle r_2 \rangle\langle x_2\rangle} \\
     - \frac{1-F_1}{\tau_1}\frac{\Cov(r_1,x_2)}{\langle r_1 \rangle\langle x_2\rangle} - \frac{1-F_2}{\tau_2}\frac{\Cov(r_2,x_1)}{\langle r_2 \rangle\langle x_1\rangle} \\
     + \frac{E_2\phi_1}{E_1\tau_2}\frac{\Cov(r_S, x_1)}{\langle r_S \rangle \langle x_1 \rangle} + \frac{E_1\phi_2}{E_2\tau_1}\frac{\Cov(r_S, x_2)}{\langle r_S \rangle \langle x_2 \rangle} \\
     - \frac{\phi_1}{\tau_2}\frac{\Cov(r_S, x_2)}{\langle r_S \rangle \langle x_2 \rangle} - \frac{\phi_2}{\tau_1}\frac{\Cov(r_S, x_1)}{\langle r_S \rangle \langle x_1 \rangle}\ .
\end{multline}
Note, to obtain \cref{eq: gbd1}, we used the fact that $\psi_1 E_2/E_1 = \psi_2, \phi_1 E_2/E_1 = \phi_2$, which follows directly from the definitions of these quantities, see \crefrange{eq: E_i appendix}{eq: psi_i appendix}. 

Since the covariance matrix must be positive semidefinite, we have the Cauchy-Schwarz inequality:
\begin{equation}
    -\CV_1\CV_2 \leq \eta_{12} \leq \CV_1\CV_2
\end{equation}
Combining this with the inequality of arithmetic and geometric means inequality gives
\begin{equation}
    \eta_{12} \geq -\frac{\CV_1^2 + \CV_2^2}{2}\ .
\end{equation}
Using this to lower bound $\eta_{12}$ in ~\cref{eq: gbd1} and collecting $\CV_1^2, \CV_2^2$ terms on the left-side yields
\begin{multline}\label{eq: gbd2}
     \left\{\frac{(1-E_1)E_2}{E_1\tau_2} + \frac{1}{2}\left[\frac{1-E_1}{\tau_1} 
     + \frac{1-E_2}{\tau_2}\right]\right\}\CV_1^2 \\
     + \left\{\frac{(1-E_2)E_1}{E_2\tau_1} + \frac{1}{2}\left[\frac{1-E_1}{\tau_1} 
     + \frac{1-E_2}{\tau_2}\right]\right\}\CV_2^2 \geq \Lambda \\
     +\frac{E_2\langle s_{11}\rangle  - E_1\langle s_{21} \rangle}{E_1\tau_2\langle x_1 \rangle} + \frac{E_1\langle s_{22}\rangle  - E_2\langle s_{12} \rangle}{E_2\tau_1\langle x_2 \rangle}\ .
\end{multline}
From the definition of the assembly efficiencies $E_1,E_2$ (\cref{eq: E_i appendix}) and Little's law (\cref{eq: Little's Law}), we have the following identities:
\begin{align}
    \frac{E_1\tau_2 \langle x_1 \rangle}{\delta_1} &= \frac{E_2\tau_1 \langle x_2 \rangle}{\delta_2} \\
     \tau_i &= \frac{1-E_i}{\beta_i}, \quad i = 1,2 \label{eq: tau_i identity}\ .
\end{align}
Thus, rescaling both sides of the inequality by $E_1\tau_2 \langle x_1 \rangle/\delta_1$, yields
\begin{multline}\label{eq: gbd3}
     \frac{E_1\tau_2}{\delta_1}\left\{\frac{(1-E_1)E_2}{E_1\tau_2} + \frac{1}{2}\left[\frac{1-E_1}{\tau_1} 
     + \frac{1-E_2}{\tau_2}\right]\right\}\frac{\CV_1^2}{1/\langle x_1 \rangle} \\
     + \frac{E_2\tau_1}{\delta_2}\left\{\frac{(1-E_2)E_1}{E_2\tau_1} + \frac{1}{2}\left[\frac{1-E_1}{\tau_1} 
     + \frac{1-E_2}{\tau_2}\right]\right\}\frac{\CV_2^2}{1/\langle x_2 \rangle} \\
     \geq \widetilde{\Lambda}
     +\frac{E_2\langle s_{11}\rangle  - E_1\langle s_{21} \rangle}{\delta_1} + \frac{E_1\langle s_{22}\rangle  - E_2\langle s_{12} \rangle}{\delta_2}\ ,
\end{multline}
where
\begin{equation}
    \widetilde{\Lambda} \coloneqq \frac{E_1\tau_2 \langle x_1 \rangle}{\delta_1}\Lambda\ .
\end{equation}
Then substituting \cref{eq: tau_i identity} leads to
\begin{multline}\label{eq: GB}
     \frac{1}{\delta_1}\left[E_2(1-E_1) + \frac{\frac{\beta_1}{\beta_2} + 1}{2}E_1(1-E_2)\right]\frac{\CV_1^2}{1/\langle x_1 \rangle} \\ 
     + \frac{1}{\delta_2}\left[E_1(1-E_2) + \frac{\frac{\beta_2}{\beta_1} + 1}{2}E_2(1-E_1)\right]\frac{\CV_2^2}{1/\langle x_2 \rangle} \\
     \geq \frac{E_2\langle s_{11}\rangle  - E_1\langle s_{21} \rangle}{\delta_1} 
     + \frac{E_1\langle s_{22}\rangle  - E_2\langle s_{12} \rangle}{\delta_2} + \widetilde{\Lambda}\ .
\end{multline}
Deriving the main results of the paper relies on constraining $\widetilde{\Lambda}$ for the various classes of systems considered in the main text, which are each a sub-class of \cref{eq: super class}. Before proceeding, we first simplify $\widetilde{\Lambda}$ using the definitions of $E_1, E_2,\phi_1,\phi_2, F_1, F_2$ (\crefrange{eq: E_i appendix}{eq: F_i appendix}) and Little's law (\cref{eq: Little's Law}). Combining these definitions with the bilinearity of covariances, we obtain the following simplified expression:
\begin{multline}\label{eq: tilde-lambda}
    \widetilde{\Lambda} = \frac{\delta_1\delta_2\langle r_A \rangle}{\langle R_1^\pm \rangle \langle R_2^\pm \rangle}\\
    \times \Cov\left(\frac{r_1 + r_S}{\delta_1} - \frac{r_2 + r_S}{\delta_2},\frac{x_1}{\delta_1} - \frac{x_2}{\delta_2}\right)\ .
\end{multline}

\section{Specialization of the general bound to restricted classes of assembly processes}
\subsection{Derivation of \cref{eq: beta asymmetry}}~\label{sect: beta appendix}
We consider the system defined by Eqs.~(\ref{eq: 2016 with co-synthesis - subunit dynamics},~\ref{eq: 2016 with co-synthesis - complexing dynamics}) from the main text, which corresponds to a restriction of the general class of systems \cref{eq: super class} where
\begin{equation}
    \begin{aligned}
        r_1(\boldsymbol{u}^\mathrm{OL},\boldsymbol{u}^\mathrm{CL}, \boldsymbol{x}) &= r_2(\boldsymbol{u}^\mathrm{OL},\boldsymbol{u}^\mathrm{CL}, \boldsymbol{x}) \\
        &\eqqcolon r_I(\boldsymbol{u}^\mathrm{OL},\boldsymbol{u}^\mathrm{CL}, \boldsymbol{x}) 
    \end{aligned}
\end{equation}
and $\delta_1 = \delta_2 = 1$. %

Under these restrictions $\frac{r_1 + r_S}{\delta_1} - \frac{r_2 + r_S}{\delta_2} = 0$, so that
\begin{equation}
    \widetilde{\Lambda} = \frac{\langle r_A \rangle}{\langle R_1^\pm \rangle \langle R_2^\pm \rangle}\Cov\left(0, \frac{x_1}{\delta_1} - \frac{x_2}{\delta_2}\right) = 0
\end{equation}

Additionally, the restrictions lead to the following symmetries in flux ratios, see Eqs.~(\ref{eq: E_i appendix},~\ref{eq: F_i appendix}):
\begin{equation}
    \begin{aligned}
        E_1 &= E_2 \eqqcolon E \\
        F_1 &= F_2 \eqqcolon F %
    \end{aligned}
\end{equation}
Then using the definition of average (co-)step-sizes (\cref{eq: co-step sizes} we have
\begin{equation}
    \begin{aligned}
        \langle s_{11} \rangle &= \langle s_{22} \rangle = 1 \\ %
        \langle s_{12} \rangle &= \langle s_{21} \rangle = (F+E)/2 \ .%
    \end{aligned}
\end{equation}
The general bound \cref{eq: GB} thus leads to 
\begin{equation}\label{eq: weighted average}
    \frac{3 + \frac{\beta_1}{\beta_2}}{2}\frac{\CV_1^2}{1/\langle x_1 \rangle} +  \frac{3 + \frac{\beta_2}{\beta_1}}{2}\frac{\CV_2^2}{1/\langle x_2 \rangle} \geq 1 + \frac{1-F}{1-E}
\end{equation}
To obtain \cref{eq: beta asymmetry} we assume without loss of generality that $\beta_2 \geq \beta_1$. Thus, 
\begin{multline}
    \frac{3 + \frac{\beta_2}{\beta_1}}{2}\left(\frac{\CV_1^2}{1/\langle x_1 \rangle} + \frac{\CV_2^2}{1/\langle x_2 \rangle}\right) \\
    \geq \frac{3 + \frac{\beta_1}{\beta_2}}{2}\frac{\CV_1^2}{1/\langle x_1 \rangle} +  \frac{3 + \frac{\beta_2}{\beta_1}}{2}\frac{\CV_2^2}{1/\langle x_2 \rangle}\ , 
\end{multline}
which combined with \cref{eq: weighted average} yields \cref{eq: beta asymmetry}, as desired.

While \cref{eq: beta asymmetry} is easily interpretable and establishes diverging fluctuations for any finite degradation rate asymmetry if $F \neq 100\%$, it can only be achieved for $\beta_1 = \beta_2$ and is increasingly loose for larger degradation rate asymmetries. For example, suppose a system has $F = 0\%, E = 90\%$, $\beta_2 = 10 \beta_1$ and is feedback controlled such that $X_1$ achieves exactly Poisson fluctuations, i.e., $\frac{\CV^2_1}{1/\langle x_1 \rangle} = 1$. \cref{eq: beta asymmetry} then tells us $\frac{\CV^2_2}{1/\langle x_2 \rangle} \gtrapprox 0.69$, thus permitting the possibility that $X_2$ fluctuations are suppressed to 69\% of Poisson levels. However, the more severe (but less easily interpreted) bound given by \cref{eq: weighted average} tells us that $\frac{\CV^2_2}{1/\langle x_2 \rangle} \gtrapprox 1.45$, meaning $X_2$ fluctuations must be at least 145\% of Poisson levels, which is more than twice the minimum value permitted by the looser bound.

\subsection{Derivation of \cref{eq: hoc bound}}\label{sect: delta generalization}
Now we consider the system \cref{eq: higher-order complexes} from the main text, which corresponds to a restriction of the general class of systems \cref{eq: super class} where 
\begin{equation}
    \begin{aligned}
    r_1(\boldsymbol{u}^\mathrm{OL},\boldsymbol{u}^\mathrm{CL}, \boldsymbol{x}) &= \lambda_1 r_I(\boldsymbol{u}^\mathrm{OL},\boldsymbol{u}^\mathrm{CL},\boldsymbol{x})\\
    r_2(\boldsymbol{u}^\mathrm{OL},\boldsymbol{u}^\mathrm{CL}, \boldsymbol{x}) &= \lambda_2 r_I(\boldsymbol{u}^\mathrm{OL},\boldsymbol{u}^\mathrm{CL}, \boldsymbol{x}) \\
    r_S(\boldsymbol{u}^\mathrm{OL},\boldsymbol{u}^\mathrm{CL}, \boldsymbol{x}) &= 0
    \end{aligned}
\end{equation}
Under these restrictions $\frac{r_1 + r_S}{\delta_1} - \frac{r_2 + r_S}{\delta_2} = \left(\frac{\lambda_1}{\delta_1} - \frac{\lambda_2}{\delta_2}\right)r_I$, so that
\begin{equation}\label{eq: lambda epsilon}
    \widetilde{\Lambda} = \varepsilon\frac{\delta_1\delta_2\langle r_A \rangle}{\langle R_1^\pm \rangle \langle R_2^\pm \rangle} \Cov\left(r_I, \frac{x_1}{\delta_1} - \frac{x_2}{\delta_2}\right) \ ,
\end{equation}
where $\varepsilon \coloneqq \frac{\lambda_1}{\delta_1} - \frac{\lambda_2}{\delta_2}$. 

Thus, if $\lambda_1 = \delta_1, \lambda_2 = \delta_2$ then $\varepsilon = 0$. This leads to $\widetilde{\Lambda} = 0$ so long as $r_I$ has finite variance, as any physical rate function should. Additionally, $\varepsilon = 0$ leads to
\begin{equation}
    \begin{aligned}
        E_1 &= E_2 \eqqcolon E \\
        F_1 &= F_2 \eqqcolon F\ ,
    \end{aligned}
\end{equation}
see Eqs.(\ref{eq: E_i appendix},~\ref{eq: F_i appendix}). Then using the definition of average (co-)step-sizes (\cref{eq: co-step sizes}) we have
\begin{equation}
\begin{aligned}
    \langle s_{11} \rangle &= 1 + \frac{(F+E)(\delta_1 - 1)}{2}\\
    \langle s_{22} \rangle &= 1 + \frac{(F+E)(\delta_2 - 1)}{2}\\
    \langle s_{21} \rangle &= \frac{(F+E)\delta_1}{2}\\
    \langle s_{12} \rangle &= \frac{(F+E)\delta_2}{2} 
\end{aligned}
\end{equation}
The general bound \cref{eq: GB} thus leads to 
\begin{equation}\label{eq: weighted average deltas}
    \frac{3 + \frac{\beta_1}{\beta_2}}{1 + \frac{\delta_1}{\delta_2}}\frac{\CV_1^2}{1/\langle x_1 \rangle} +  \frac{3 + \frac{\beta_2}{\beta_1}}{1 + \frac{\delta_2}{\delta_1}}\frac{\CV_2^2}{1/\langle x_2 \rangle} \geq 1 + \frac{1-F}{1-E}\ .
\end{equation}
Without loss of generality $\beta_2 \geq \beta_1$ leads to the following (loose) bound:
\begin{multline}
    \frac{1}{1 + \frac{\delta_1}{\delta_2}}\frac{\CV_1^2}{1/\langle x_1 \rangle} +  \frac{1}{1 + \frac{\delta_2}{\delta_1}}\frac{\CV_2^2}{1/\langle x_2 \rangle} \\
    \geq \frac{1}{3 + \frac{\beta_2}{\beta_1}}\left(1 + \frac{1-F}{1-E}\right)\ .
\end{multline}
Upper bounding the left side by substituting $\min\left\{\delta_1/\delta_2,\delta_2/\delta_1\right\}$ for both delta ratios leads to \cref{eq: hoc bound}, as desired. 

\subsection{Derivation of \cref{eq: dual input OL bound}}\label{sect: OL appendix}
In the main text we present the bound \cref{eq: dual input OL bound} for the class of systems defined by restricting general class of systems \cref{eq: super class} such that 
\begin{equation}\label{eq: OL r_s}
    r_S(\boldsymbol{u}^\mathrm{OL},\boldsymbol{u}^\mathrm{CL}, \boldsymbol{x}) = r_S(\boldsymbol{u}^\mathrm{OL})
\end{equation}
and
\begin{equation}\label{eq: ol restrictions}
    \begin{aligned}
        r_1(\boldsymbol{u}^\mathrm{OL},\boldsymbol{u}^\mathrm{CL}, \boldsymbol{x}) &= r_1(\boldsymbol{u}^\mathrm{OL})\\
        r_2(\boldsymbol{u}^\mathrm{OL},\boldsymbol{u}^\mathrm{CL}, \boldsymbol{x}) &= r_2(\boldsymbol{u}^\mathrm{OL})
    \end{aligned}
\end{equation}
Here we consider a more general class of systems where for each $i = 1,2$
\begin{multline}\label{eq: unequal synthesis rates appendix}
r_i(\boldsymbol{u}^\mathrm{OL},\boldsymbol{u}^\mathrm{CL}, \boldsymbol{x}) = \lambda_i r_I(\boldsymbol{u}^\mathrm{OL},\boldsymbol{u}^\mathrm{CL}, \boldsymbol{x}) \\ + \Delta_i(\boldsymbol{u}^\mathrm{OL}) \ , 
\end{multline}
In the limit that $\lambda_1 = \lambda_2 = 0$, we recover the system defined by the restrictions \cref{eq: ol restrictions}.

Under the restrictions Eqs.~(\ref{eq: OL r_s},~\ref{eq: unequal synthesis rates appendix}) we have 
\begin{equation}
\begin{aligned}\label{eq: dual input appendix step-sizes}
    \langle s_{11} \rangle &= \frac{1-F_1 + \phi_1 + 1-E_1 + \delta_1 E_1 + \delta_1 \psi_1}{2}\\ %
    \langle s_{22} \rangle &= \frac{1-F_2 + \phi_2 + 1-E_2 + \delta_2 E_2 + \delta_2 \psi_2}{2}\\ %
    \langle s_{21} \rangle &= \frac{\phi_2 + \delta_1 E_2 + \delta_1 \psi_2}{2}\\ %
    \langle s_{12} \rangle &= \frac{\phi_1 + \delta_2 E_1 + \delta_2 \psi_1}{2}\\ %
\end{aligned}
\end{equation}
see Eqs.~(\ref{eq: co-step sizes},~\ref{eq: E_i appendix}-\ref{eq: F_i appendix}).

We wish to show \cref{eq: dual input OL bound} holds for this system. To do this, we first consider the time evolution of the ensemble averaged subunit abundances conditioned on the history of the open-loop variables $\boldsymbol{u}^\mathrm{OL}$. When conditioning on a particular history $\boldsymbol{u}^\mathrm{OL}[0,t]$, the time evolution of the conditional distribution follows a master equation (\cref{eq: cCME}) where $\boldsymbol{u}^\mathrm{OL}$ can be regarded as a deterministic time-dependent signal. Thus, we have the usual moment evolution equations (cf. \cref{eq: mean evolution}) for the conditional moments:
\begin{align}
    \frac{\mathrm{d}\overline{x_1}}{\mathrm{d}t} &= \overline{r_1} + \overline{r_S} + \delta_1(\overline{r_D} - \overline{r_A}) - \beta_1 \overline{x_1} \label{eq: x1-bar} \\ 
    \frac{\mathrm{d}\overline{x_2}}{\mathrm{d}t} &= \overline{r_2} + \overline{r_S} + \delta_2(\overline{r_D} - \overline{r_A}) - \beta_2 \overline{x_2} \ ,\label{eq: x2-bar} 
\end{align}
\comm{where for any dynamic random variable $y$, we define $\overline{x}(t) \coloneqq \left\langle x | \boldsymbol{u}^\mathrm{OL}[0,t]\right\rangle(t)$ to represent the time-dependent conditional ensemble average given a particular history of the open-loop control variables. Note, in Eqs.~(\ref{eq: x1-bar},~\ref{eq: x2-bar}) and below we suppress the time-dependence of these conditional averages for notational simplicity.} Rescaling Eqs.~(\ref{eq: x1-bar},~\ref{eq: x2-bar}) by $1/\delta_1, 1/\delta_2$, respectively, and subtracting leads to
\begin{multline}\label{eq: d-diff}
     \frac{\mathrm{d}\left(\frac{\overline{x_1}}{\delta_1} - \frac{\overline{x_2}}{\delta_2}\right)}{\mathrm{d}t} = \frac{\overline{r_1} + \overline{r_S}}{\delta_1} - \frac{\overline{r_2} + \overline{r_S}}{\delta_2}\\
     - \frac{\beta_1 \overline{x_1}}{\delta_1} + \frac{\beta_2 \overline{x_2}}{\delta_2}\ .
\end{multline}
Multiplying both sides of \cref{eq: d-diff} by $\frac{\overline{x_1}}{\delta_1} - \frac{\overline{x_2}}{\delta_2}$ and time averaging leads to %
\begin{multline}\label{eq: diff^2}
    \frac{1}{2}\left\langle \frac{\mathrm{d}\left(\frac{\overline{x_1}}{\delta_1} - \frac{\overline{x_2}}{\delta_2}\right)^2}{\mathrm{d}t} \right\rangle^* \\
    = \Cov^*\left(\frac{\overline{r_1}+\overline{r_S}}{\delta_1} - \frac{\overline{r_2}+\overline{r_S}}{\delta_2},\frac{\overline{x_1}}{\delta_1} - \frac{\overline{x_2}}{\delta_2}\right) \\
    + \left\langle \frac{\overline{r_1}+\overline{r_S}}{\delta_1} - \frac{\overline{r_2}+\overline{r_S}}{\delta_2} \right\rangle^*\left\langle \frac{\overline{x_1}}{\delta_1} - \frac{\overline{x_2}}{\delta_2} \right\rangle^* \\
    - \Cov^*\left(\frac{\beta_1 \overline{x_1}}{\delta_1} - \frac{\beta_2 \overline{x_2}}{\delta_2}, \frac{\overline{x_1}}{\delta_1} - \frac{\overline{x_2}}{\delta_2} \right) \\
    - \left\langle \frac{\beta_1 \overline{x_1}}{\delta_1} - \frac{\beta_2 \overline{x_2}}{\delta_2} \right\rangle^*\left\langle \frac{\overline{x_1}}{\delta_1} - \frac{\overline{x_2}}{\delta_2} \right\rangle^*\ ,
\end{multline}
where $\Cov^*(\cdot,\cdot)$ denotes the temporal covariance defined through the long-time average in \cref{eq: long-time averaging}. 

The left side of \cref{eq: diff^2} must vanish at stationarity, otherwise one of $\overline{x_1},\overline{x_2}$ must go to $\infty$ as $t\to \infty$. Additionally, at stationarity we have
\begin{equation}
    0 = \left\langle \frac{\mathrm{d}\left(\frac{\overline{x_1}}{\delta_1} - \frac{\overline{x_2}}{\delta_2}\right)}{\mathrm{d}t} \right\rangle^*\ ,
\end{equation}
which implies  
\begin{multline}
    \left\langle \frac{\overline{r_1}+\overline{r_S}}{\delta_1} - \frac{\overline{r_2}+\overline{r_S}}{\delta_2} \right\rangle^* \\
    = \left\langle \frac{\beta_1\overline{x_1}}{\delta_1} - \frac{\beta_2\overline{x_2}}{\delta_2} \right\rangle^*
\end{multline}
Combining this with \cref{eq: diff^2} and using the bilinearity of covariance, we have
\begin{multline}\label{eq: cov-bar identity}
    \Cov^*\left(\frac{\overline{r_1}+\overline{r_S}}{\delta_1} - \frac{\overline{r_2}+\overline{r_S}}{\delta_2},\frac{\overline{x_1}}{\delta_1} - \frac{\overline{x_2}}{\delta_2}\right)
    \\=\frac{\beta_1}{\delta_1^2}\Var^*(\overline{x_1}) + \frac{\beta_2}{\delta_2^2}\Var^*(\overline{x_2})
    \\- \frac{\beta_1+\beta_2}{\delta_1\delta_2}\Cov^*(\overline{x_1},\overline{x_2})\ .
\end{multline}

The basic identity for the variance of a weighted sum of random variables gives
\begin{multline}
    \frac{\beta_1}{\delta_1^2}\Var^*\left(\overline{x_1}\right) + \frac{\beta_2}{\delta_2^2}\Var^*\left(\overline{x_2}\right) =  2\frac{\sqrt{\beta_1\beta_2}}{\delta_1\delta_2}\Cov^*(\overline{x_1},\overline{x_2}) \\
    + \Var^*\left(\frac{\sqrt{\beta_1}\overline{x_1}}{\delta_1} - \frac{\sqrt{\beta_2}\overline{x_2}}{\delta_2} \right)\ .
\end{multline}
Along with \cref{eq: cov-bar identity} and the non-negativity of variances we thus have
\begin{multline}\label{eq: cov-bar ineq}
    \Cov^*\left(\frac{\overline{r_1}+\overline{r_S}}{\delta_1} - \frac{\overline{r_2}+\overline{r_S}}{\delta_2},\frac{\overline{x_1}}{\delta_1} - \frac{\overline{x_2}}{\delta_2}\right) \\
    \geq \frac{2\left(\sqrt{\beta_1\beta_2} - \frac{\beta_1 + \beta_2}{2}\right)}{\delta_1\delta_2}\Cov^*(\overline{x_1},\overline{x_2})\ .
\end{multline}
Note, $\sqrt{\beta_1\beta_2} - \frac{\beta_1 + \beta_2}{2} \leq 0$ by the AM-GM inequality. 
As in Sect.~\ref{sect: delta generalization}, let us define $\varepsilon \coloneqq \frac{\lambda_1}{\delta_1} - \frac{\lambda_2}{\delta_2}$. Then by bilinearity of covariance and the definition of the subunit synthesis rates (\cref{eq: unequal synthesis rates appendix}), we have
\begin{multline}\label{eq: cu cov-bar1}
    \Cov^*\left(\frac{\overline{r_1}+\overline{r_S}}{\delta_1} - \frac{\overline{r_2}+\overline{r_S}}{\delta_2},\frac{\overline{x_1}}{\delta_1} - \frac{\overline{x_2}}{\delta_2}\right) \\
    = \Cov\left(\frac{\overline{\Delta_1}+\overline{r_S}}{\delta_1} - \frac{\overline{\Delta_2}+\overline{r_S}}{\delta_2},\frac{\overline{x_1}}{\delta_1} - \frac{\overline{x_2}}{\delta_2}\right) \\
    + \epsilon \Cov\left(\overline{r_I},\frac{\overline{x_1}}{\delta_1} - \frac{\overline{x_2}}{\delta_2}\right)
\end{multline}
Using ergodicity and the fact that $\Delta_1,\Delta_2,r_S$ depend only on the conditioned history, we have
\begin{multline}\label{eq: cu cov-bar2}
    \Cov^*\left(\frac{\overline{\Delta_1}+\overline{r_S}}{\delta_1} - \frac{\overline{\Delta_2}+\overline{r_S}}{\delta_2},\frac{\overline{x_1}}{\delta_1} - \frac{\overline{x_2}}{\delta_2}\right)\\
    = \Cov\left(\frac{\Delta_1+r_S}{\delta_1} - \frac{\Delta_2+r_S}{\delta_2},\frac{x_1}{\delta_1} - \frac{x_1}{\delta_2}\right)\ .
\end{multline}
Eqs.~(\ref{eq: tilde-lambda}),~(\ref{eq: cov-bar ineq})-(\ref{eq: cu cov-bar2}) and bilinearity of covariance then gives 
\begin{multline}
    \frac{\langle R_1^\pm \rangle \langle R_2^\pm \rangle}{\delta_1\delta_2\langle r_A \rangle}\widetilde{\Lambda} \\
    \geq \frac{2}{\delta_1\delta_2}\left(\sqrt{\beta_1\beta_2} - \frac{\beta_1 + \beta_2}{2}\right)\Cov(\overline{x_1},\overline{x_2}) \\
    + \epsilon\Cov\left(r_I,\frac{x_1}{\delta_1} - \frac{x_2}{\delta_2}\right) \\ -\epsilon\Cov\left(\overline{r_I},\frac{\overline{x_1}}{\delta_1} - \frac{\overline{x_2}}{\delta_2}\right)  
\end{multline}
The Cauchy-Schwarz inequality implies that the covariance terms involving $r_I$ vanish for $|\varepsilon| \to 0$ so long as $r_I$ has finite variance. Thus, any conclusion obtained for $\varepsilon = 0$ is asymptotically valid for small $|\varepsilon|$. Taking $\varepsilon = 0$ we have
\begin{align}
     \frac{\langle R_1^\pm \rangle \langle R_2^\pm \rangle}{\langle r_A \rangle}\widetilde{\Lambda}\notag&\geq 2\left(\sqrt{\beta_1\beta_2} - \frac{\beta_1 + \beta_2}{2}\right)\\
     &\hspace{1.5cm} \times \Cov(\overline{x_1},\overline{x_2})\\
     &\geq 2\left(\sqrt{\beta_1\beta_2} - \frac{\beta_1 + \beta_2}{2}\right)\notag\\
     &\hspace{1.5cm} \times\sqrt{\Var(\overline{x_1})\Var(\overline{x_2})}\label{eq: CS} \\
     &\geq 2\left(\sqrt{\beta_1\beta_2} - \frac{\beta_1 + \beta_2}{2}\right)\notag\\
     &\hspace{1.5cm} \times\sqrt{\Var(x_1)\Var(x_2)}\label{eq: LOTV}\ ,
\end{align}
where \cref{eq: CS} follows from the Cauchy-Schwarz inequality and \cref{eq: LOTV} follows from the law of total variance. Little's law (\cref{eq: Little's Law}) and the flux balance relations (\cref{eq: R_i}) give
\begin{equation}
    \frac{\langle r_A \rangle}{\langle R_1^\pm \rangle \langle R_2^\pm \rangle} = \frac{\langle r_A\rangle \tau_1\tau_2}{\langle x_1 \rangle \langle x_2 \rangle}\ .
\end{equation}
Thus,
\begin{align}
    \widetilde{\Lambda} \notag&\geq 2\langle r_A \rangle\tau_1\tau_2\\
     &\hspace{1cm} \times\left(\sqrt{\beta_1\beta_2} - \frac{\beta_1 + \beta_2}{2}\right)\CV_1\CV_2 \\
     \notag&\geq \langle r_A \rangle\tau_1\tau_2\\
     &\hspace{1cm} \times\left(\sqrt{\beta_1\beta_2} - \frac{\beta_1 + \beta_2}{2}\right)\left(\CV_1^2 + \CV_1^2\right)\ ,\label{eq: amgm1}
\end{align}
where \cref{eq: amgm1} follows from the AM-GM inequality. Again, invoking the definition of $E_1,E_2$ (\cref{eq: E_i appendix}), Little's law (\cref{eq: Little's Law}) and the flux balance relations (\cref{eq: R_i}), we have, for example, 
\begin{equation}
    \langle r_A \rangle\tau_1\tau_2\CV_1^2 = \frac{E_1}{\delta_1}\frac{1-E_2}{\beta_2}\frac{\CV_1^2}{1/\langle x_1 \rangle}\ .
\end{equation}
Combining this with the analogous expression for $\langle r_A \rangle\tau_1\tau_2\CV_2^2$, we have
\begin{multline}
    \widetilde{\Lambda} \geq \left(\sqrt{\beta_1\beta_2} - \frac{\beta_1 + \beta_2}{2}\right)\\
    \times \left(\frac{E_1}{\delta_1}\frac{1-E_2}{\beta_2}\frac{\CV_1^2}{1/\langle x_1 \rangle} + \frac{E_2}{\delta_2}\frac{1-E_1}{\beta_1}\frac{\CV_2^2}{1/\langle x_2 \rangle}\right)
\end{multline}
Substituting this inequality and the (co-)step-sizes (\cref{eq: dual input appendix step-sizes}) in \cref{eq: GB} and collecting terms gives
\begin{multline}
     \frac{1}{\delta_1}H\left(E_1,E_2,\frac{\beta_1}{\beta_2}\right)\frac{\CV_1^2}{1/\langle x_1 \rangle} \\
     + \frac{1}{\delta_2}H\left(E_2,E_1,\frac{\beta_2}{\beta_1}\right)\frac{\CV_2^2}{1/\langle x_2 \rangle} \\
     \geq \frac{E_2(2 - E_1 - F_1)}{2\delta_1} + \frac{E_1(2 - E_2 - F_2)}{2\delta_2} \ ,
\end{multline}
where
\begin{multline}
    H\left(E_1,E_2,\frac{\beta_1}{\beta_2}\right) \coloneqq E_2(1-E_1) \\
    + \left(\frac{\beta_1}{\beta_2} + 1 - \sqrt{\frac{\beta_1}{\beta_2}}\right)E_1(1-E_2) \ .
\end{multline}
It can be verified that $H\left(E_2,E_1,\beta_2/\beta_1\right) \geq H\left(E_1,E_2,\beta_1/\beta_2\right)$ for all permissible values of $E_1,E_2$ if $\beta_2/\beta_1 \geq 1$. Assuming $\beta_2 \geq \beta_1$ without loss of generality then leads to the following loose but more easily interpreted bound:
\begin{multline}\label{eq: pre-OL bound}
     \frac{1}{\delta_1}\frac{\CV_1^2}{1/\langle x_1 \rangle} + \frac{1}{\delta_2}\frac{\CV_2^2}{1/\langle x_2 \rangle} \geq \\
     \frac{1}{2}\frac{E_2(2 - E_1 - F_1)\delta_2^{-1}+ E_1(2 - E_2 - F_2)\delta_1^{-1}}{E_1(1-E_2) + \left(\frac{\beta_2}{\beta_1} + 1 - \sqrt{\frac{\beta_2}{\beta_1}}\right)E_2(1-E_1)}\ ,
\end{multline}
\cref{eq: dual input OL bound} follows by dividing \cref{eq: pre-OL bound} by $2(\delta_1^{-1}+\delta_2^{-1})$ and upper bounding the left side by substituting $\min\left\{\delta_1/\delta_2,\delta_2/\delta_1\right\}$ for both delta ratios. 

\section{\comm{Analytical treatment of singular limits}}
\comm{
Next, we consider systems defined by Eqs.~(\ref{eq: 2016 with co-synthesis - subunit dynamics},~\ref{eq: 2016 with co-synthesis - complexing dynamics}) of the main text and analyze their behaviour in two singular limits.
\subsection{\comm{$E=F=1$}}\label{sect: E=F=1}
\comm{If $E = 1$ then $\beta_1 \langle x_1 \rangle = \beta_2 \langle x_2 \rangle = 0$ (see Eqs.~(\ref{eq: E defn},~\ref{eq: flux balance})). Thus the fraction of time that the individual degradation rate for either component is non-zero has measure $0$ and we can regard these rates as $0$. Similarly, if $F = 1$ then $r_I = 0$ (see Eq.~\eqref{eq: F defn}). Thus, $X_1$ and $X_2$ molecules always enter and leave the system as a pair. Thus, $x_2 = x_1 + \omega$ at all times, where $\omega$ is the initial offset between $x_2$ and $x_1$, which illustrates the loss of ergodicity due to the dependence of the long-term behaviour on initial conditions. Because $x_2 - x_1$ is constant, the two-component dynamics reduce to the following one-variable problem:}
\begin{equation}
    x_1 \overset{r_S}{\xrightarrow{\hspace*{1.1cm}}} x_1 + 1 \qquad x_1 \overset{r_A}{\xrightarrow{\hspace*{1.1cm}}} x_1 - 1\ ,
\end{equation}
where any functional dependence of $r_S, r_A$ on $X_2$ levels is replaced with the identity $x_2 = x_1 + \omega$. Now, as in Ref.~\cite{Laurenti2018}, we take $\omega = 0$ and let $r_S = \alpha x_0$, where $x_0$ follows a Poisson process, and $r_A = \gamma x_1 x_2 = \gamma x_1(x_1 + \omega)$. 
While the stationary moment equations are not exactly solvable due to the nonlinearity of $r_A$, we can follow a linear noise approximation ~\cite{vank1992} to approximately determine the fluctuations
\begin{equation}
    \eta_{11} \approx \frac{1}{2} \left(\frac{1}{\left\langle x_1\right\rangle } + \frac{\tau _0}{\left(2 \tau _0+\tau _1\right) \left\langle x_0\right\rangle }\right)\ .
\end{equation}
In this approximation, the quadratic degradation rate can reduce $\eta_{11}$ to $1/2$ of Poisson fluctuations as long as $X_0$ fluctuates on much faster time scales than $X_1$ -- i.e. $\tau_0 \ll \tau_1$. The noise-reducing properties of quadratic degradation terms has been discussed previously~\cite{Nicolis1972, Mazo1975}. The problem of regulating the fluctuations of single-variable dynamics through self-enhanced degradation is fundamentally different from the problem of noise control in two-component assembly processes where each component's abundance can individually fluctuate stochastically, i.e. where one component's abundance is not exactly determined by the other component's abundance.}

\subsection{\comm{$\beta_2/\beta_1 = \infty$, $E = 1$}}\label{sect: beta2/beta1 = infty, E=1}
\comm{We here consider $\beta_1 = 0$ and $\beta_2>0$ along with $E = 1$.  Assuming the system reaches a stationary state implies that $\beta_2 \langle x_2 \rangle = 0$ which follows from the flux balance relation (Eq.~\eqref{eq: flux balance}) and the definition of $E$ (Eq.~\eqref{eq: E defn}). Thus, for $\beta_2 > 0$ we have $\langle x_2 \rangle = 0$. This implies that the fraction of time that $x_2$ is non-zero has measure $0$, and we can regard $x_2 = 0$. This implies that systems with $\beta_1 = 0,\ \beta_2 >0$ can only achieve $E=1$ if the assembly rate is effectively $\infty$ for non-zero $x_2$ such that $X_2$ molecules are immediately sequestered to the complex upon synthesis. The corresponding $X_1$-dynamics then become %
\begin{equation}
    x_1 \overset{r_I}{\xrightarrow{\hspace*{1.1cm}}} x_1 + 1 \qquad x_1 \overset{r_I}{\xrightarrow{\hspace*{1.1cm}}} x_1 - 1\ ,
\end{equation}
which corresponds to an unbiased random walk on $\mathbb{Z}$ that does not reach a stationary state, as the variance of $X_1$ diverges.}%

\section{Numerical simulation details}\label{sect: numerics}
Stochastic realizations of the presented processes were sampled from the CME using the standard stochastic simulation algorithm~\cite{doob1945, Gillespie1977} implemented in \CC11. Analysis and plotting of numerical data were performed using scripts implemented in MATLAB R2023b. To estimate stationary temporal statistics, three independent simulations initialized with different pseudo-random number generator seeds were simulated until each reaction defining the system occurred at least $10^7$ times. We verified each simulation satisfied the first- and second-order moment invariants for stationary processes within 2\% and 3\% relative error, respectively, see Eqs.~(\ref{eq: flux balance},~\ref{eq: covariance balance}). The stationary temporal means and Poisson-normalized squared-CVs are plotted as the mean across these three simulations for each system in Fig.~\ref{fig: dual-input CL}. The standard error relative to the estimated value for both the numerically estimated means and Poisson-normalized squared-CVs was at most $\sim 0.1\%$ %
across all three systems. Time-dependent ensemble statistics were estimated across 8000 independent simulations initialized with different pseudo-random number generator seeds. Time was discretized with $\Delta t = 0.1$ to generate the ensemble average data plotted in Fig.~\ref{fig: dual-input CL}. 

The assembly process parameters were held at $\gamma = 100, \ \lambda = 1, \ \beta_3 = 1$ for each of the three example systems in Fig~\ref{fig: dual-input CL}, see ~\cref{eq: assembly dynamics}. The additional parameters for the antithetic integral feedback system (Fig.~\ref{fig: dual-input CL}(a)) were set at $\mu = 50, \ \theta = 3, \ k = 2, \ \beta_4 = 0.5$, see Eqs.~(\ref{eq: AIF module-1},~\ref{eq: AIF module-2}). The additional parameters for the symmetric direct feedback system (Fig.~\ref{fig: dual-input CL}(b)) were set at $\nu = 1000, \ n = -4, \ \kappa = 1$, see ~\cref{eq: symmetric rates}. Finally, the additional parameters for the asymmetric direct feedback system (Fig.~\ref{fig: dual-input CL}(c)) were set at $\nu_1 = 1200, \ n_1 = -3, \ \kappa_1 = 1, \ \nu_2 = 900, \ n_2 = 2, \ \kappa_2 = 1$, see ~\cref{eq: asymmetric rates}

\bibliography{bibliography}

\end{document}